\documentclass[twocolumn]{aastex631}

\usepackage{mathtools}
\usepackage{amsmath}
\usepackage{booktabs}
\usepackage{graphicx}
\usepackage{multirow}
\usepackage{natbib}
\usepackage{CJKutf8}

\graphicspath{{./}{figures/}}

\begin{document}

\title{Interaction of an outflow with surrounding gaseous clouds as the origin of the late-time radio flares in TDEs}

\author{Jialun Zhuang \begin{CJK}{UTF8}{gbsn}(庄嘉伦)\end{CJK} } 
\author{Rong-Feng Shen  \begin{CJK}{UTF8}{gbsn}(申荣锋)\end{CJK} }
\affiliation{School of Physics and Astronomy, Sun Yat-Sen University, Zhuhai, 519000,  China }
\affiliation{CSST Science Center for the Guangdong-Hongkong-Macau Greater Bay Area, Sun Yat-Sen University, Zhuhai, 519082, China}

\author{Guobin Mou \begin{CJK}{UTF8}{gbsn}(牟国斌)\end{CJK}}
\affiliation{Department of Physics and Institute of Theoretical Physics, Nanjing Normal University, Nanjing 210023, China}

\author{Wenbin Lu \begin{CJK}{UTF8}{gbsn}(鲁文宾)\end{CJK}}
\affiliation{Department of Astronomy, University of California, Berkeley, CA 94720-3411, USA}

\email{E-mail: zhuangjlun@mail2.sysu.edu.cn (JZ); shenrf3@mail.sysu.edu.cn (RS); gbmou@njnu.edu.cn (GM); wenbinlu@berkeley.edu (WL)}

\begin{abstract}
Close encounter between a star and a supermassive black hole (SMBH) results in the tidal disruption of the star, known as a tidal disruption event (TDE).  
Recently, a few TDEs, e.g., ASASSN-15oi and AT2018hyz, have shown late-time (hundreds of days after their UV/optical peaks) radio flares with radio luminosities of $10^{38\sim39}$ erg/s. 
The super-Eddington fallback or accretion in a TDE may generate a mass outflow.
Here we investigate a scenario that the late-time radio flares come from the interaction of the outflow with the circum-nuclear gaseous clouds, in addition to the slow-evolving emission component due to the outflow-diffuse medium interaction.  
We calculate the associated radio temporal and spectral signatures and find that they reproduce well the observations.
The outflows have the inferred velocity of 0.2$c\sim0.6$$c$, the total mass of $10^{-3}\sim10^{-1}$ $\mathrm{M_{\odot}}$ and the ejection duration of a month to a year.
The distances of the clouds to the SMBH are $0.1\sim1$ pc. 
This scenario has advantages in explaining the long delay, sharpness of the rise and the multiplicity of the late radio flares.
Future observations may build up a much larger sample of late-time radio flares and enable their use as a probe of the TDE physics and the host circumnuclear environment.
\end{abstract}



\section{Introduction} \label{sec:intro}

Most galaxies are considered to contain a supermassive black hole (BH) with a mass above $\sim$ $10^{6}$ $\mathrm{M_{\odot}}$ at its nuclei \citep{2005SSRv..116..523F}.
A star wandering too close to the supermassive BH would be tidally disrupted as a tidal disruption event (TDE) \citep{1988Natur.333..523R,1989IAUS..136..543P}.
Accretion of the stellar debris would produce a bright flare in X-ray or UV/optical, which would last for months to years with the light curve decaying as $t^{-\frac{5}{3}}$ \citep{1988Natur.333..523R,1989IAUS..136..543P}.

In TDEs, outflow can be produced via various mechanisms.
At the moment of disruption, the unbound debris would fly away at a velocity of $10^{4}$  km/s  \citep{2016ApJ...822...48G,2019MNRAS.487.4083Y}.
Besides, a fraction of the matter initially bound to the BH is likely to be blown away with a velocity of $\sim$ 0.1$c$, where $c$ is the light speed, by the radiation pressure when the fallback rate, or the accretion rate, of these bound debris is super-Eddington at early times \citep{2009MNRAS.400.2070S,2023MNRAS.523.4136B}.
Furthermore, for BHs with masses greater than $10^{7}$ $\mathrm{M_{\odot}}$,  the self-crossing of the fallback stream, due to relativistic apsidal precession, can also generate energetic, fast outflow, and the speed of these outflow ranges from $\sim0.01c$ to $\sim0.1c$ \citep{2020MNRAS.492..686L}.
Another possibility concerns a relativistic jet which can be generated in the presence of a large magnetic flux threading a rapidly spinning BH \citep{1977MNRAS.179..433B,2011MNRAS.418L..79T,2011Sci...333..203B,2011Natur.476..421B}.

When the high-speed outflow or jet interacts with the circumnuclear medium (CNM), it would drive a forward shock in the CNM, and produce radio emission via the synchrotron radiation \citep{1998ApJ...499..810C,2013ApJ...772...78B}.
In the past decade, dozens of TDEs indeed have shown these GRB-afterglow-like radio emissions, and their radio luminosities span from $10^{36}$ to $10^{42}$ erg/s \citep{2020SSRv..216...81A}.

However, recent TDEs radio observations reveal flares only seen at late times ($10^{2\sim3}$ days after TDE discovery). Due to their steep rise of light curves, two typical cases being ASASSN-15oi \citep{2021NatAs...5..491H} and AT2018hyz \citep{2022ApJ...938...28C}, they are difficult to be explained by the conventional outflow-CNM scenario.
The steepest evolution from the outflow-CNM interaction is $f_{\nu}\propto t^{3}$ \citep{2012MNRAS.420.3528M,2016ApJ...827..127K,2020SSRv..216...81A},  but  ASASSN-15oi and AT2018hyz, show a jump in flux density from non-detection to detection that requires a temporal power-law steeper than $t^{4}$.
Similar ones were observed in  IGR J12580+0134 \citep{2022ApJ...925..143P}, AT2019azh \citep{2022ApJ...933..176S,2022MNRAS.511.5328G} and AT2020vwl \cite{2023ATel16165....1G,2023MNRAS.tmp.1199G,2024arXiv241018665G}.

Various models have been proposed to explain these late and steeply rising radio flares in TDEs. \cite{2023arXiv230813595C} conclude that the delayed outflows, such as from delayed disk formation, are a preferred explanation.
However, early X-ray detection indicates that the accretion disk  is already present early on \citep{2023arXiv230813019G}. So the outflow is expected to be launched at an early time without a  time delay.
 Likewise, early radio detection such as AT2019azh suggests that the outflow tends to be launched at the occurrence of TDEs.

\cite{2023ApJ...957L...9T} propose a misaligned precessing jet, which is initially choked by the disk wind but later breaks out, when the disk eventually aligns itself with the BH spin axis. 
This jet may produce delayed radio rebrightening seen in  TDEs.
An alternative model is that of a decelerated off-axis jet \citep{2023MNRAS.522.4565M,2024MNRAS.527.7672S}, which we will address in Section 5.

Besides, \cite{2024arXiv240415966M} find that a CNM density profile that initially drops with radius in a power law but is followed by a constant outside the Bondi radius can explain the radio flares in  AT2020vwl, AT2019dsg, PS16dtm and ASASSN14ae.
However,  this model is difficult to explain a rise steeper than $\propto t^{3}$, like those in AT2018hyz and ASASSN-15oi, unless the external density profile rises again.
\cite{2024arXiv241020127L} studied the collision of the unbound half of the stellar debris stream with an outer dusty torus. They show that it may produce a delayed ($\sim$ years) radio flare with a rapid rise.


If discrete gaseous clouds exist in the circum-nuclear surroundings (see below), the TDE outflow will sweep across them.
\cite{2022MNRAS.510.3650M} studied the outflow–cloud interaction and show that it can generate considerable radio emission  months or years later after the TDE,  and can explain the temporal evolution of the peak frequencies in AT2019dsg, ASASSN-14li, and CSS161010.
Based on numerical simulations of the disk wind launching, \cite{2023MNRAS.521.4180B} further studied this model.
Besides the delayed radio flare, the outflow-cloud interaction may have multi-messenger signatures including neutrinos,  gamma-ray and X-ray \citep{2021MNRAS.507.1684M,2021ApJ...908..197M,2022MNRAS.514.4406W,2023MNRAS.518.5163C}.
%
%

\begin{figure}[htp]
\centering
\includegraphics[width=0.48\textwidth]{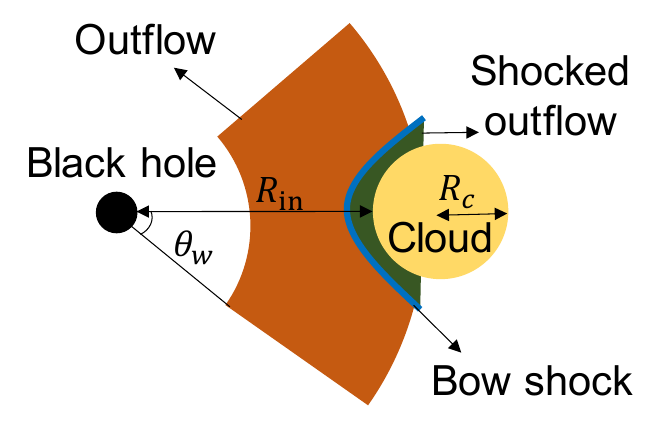}
\caption{Schematic of the interaction of the TDE outflow with the cloud. Outflow is generated at the occurrence of TDEs with a half opening angle $\theta_{w}$ and a velocity $v_{w}$.
The cloud is located at a distance of $R_{in}$ with a radius $R_{c}$.
The interaction of the outflow with the cloud leads to the formation of a bow shock, producing a late-time radio flare years after TDE. 
This illustration is not to scale.
\label{fig:structure}}
\end{figure}

Here,  we investigate the outflow-cloud interaction model for these late-time radio flares.
This collision forms a bow shock in front of the inner face of the cloud, across which electrons are accelerated and the magnetic field is amplified.
The late-time radio synchrotron emission is expected from the shocked outflow.
An illustration of the interaction of the TDE outflow with the cloud is shown in Figure \ref{fig:structure}.

In section 2, we introduce the semi-analytical model for the wind-cloud interaction and calculate the associated synchrotron emission.
We take the effect of the interaction of the outflow with a diffuse CNM  into account in section 3.
The application of our model to a few TDEs is presented in section 4.
Discussion and conclusion are given in sections 5 and 6, respectively.

\section{The outflow-cloud interaction}\label{sec:model}

\subsection{The clouds}


Observation of the Milky Way galactic center reveals tens of clouds near the Sgr $A^{\star}$, and the distribution of these clouds indicates a ring-like structure  \citep{1996A&ARv...7..289M,2005ApJ...622..346C}.
These clouds are at distances of $0.5\sim2$ pc with gas number densities of $10^{6}\sim10^{8}$ $\mathrm{cm^{-3}}$, and individual  masses of $10^{4\sim5}$ $\mathrm{M_{\odot}}$ (hence  an estimated radius of $\sim0.1$ pc)\citep{1996A&ARv...7..289M,2005ApJ...622..346C}.

As for AGN, analysis  shows that the clouds have smaller  sizes of $10^{14\sim15}$ cm but higher densities of $10^{8\sim9}$ $\mathrm{cm^{-3}}$, at similar distances of $0.01\sim0.1$ pc \citep{2014MNRAS.439.1403M, 2022MNRAS.514.1535A}.
The estimated cloud masses are $\sim10^{-6\sim-2}$ $\mathrm{M_{\odot}}$, which are much lower than those in Sgr $A^{\star}$.
We  are not aware of any physical reason for the large discrepancy in the cloud masses or whether it is just an observational bias.
Here we consider the simplest case where the outflow interacts with a cloud at a distance $R_{in}$ $\approx0.1\sim1$ pc  and of a radius $R_c$ $\approx0.01\sim0.1$ pc, like those in Sgr $A^{\star}$.


\subsection{The outflow}\label{sec_outflow}

After the disruption, the stellar debris will fall back to the vicinity of the BH after a so-called fall back timescale $t_{fb} = 130  R_{\star,0}^{3/2} M_{\star,0}^{-1} M_{bh,7}^{1/2}$   [day] \citep{2013ApJ...767...25G}, and then form an accretion disk in a so-called circularization process which lasts for $\sim t_{fb}$ or longer.
The outflow is  launched  when the BH is in the super-Eddington accretion phase.
Meanwhile, the X-ray emission from the inner accretion flow is reprocessed by the outflow into optical emission, giving rise to the TDE's optical flare (e.g., \citeauthor{2018ApJ...859L..20D} \citeyear{2018ApJ...859L..20D}).
Here we set the launching time of the accretion-driven outflow to be $t=0$.

According to the simulation of the super-Eddington accretion in TDEs, the outflow is nearly spherical but largely anisotropic \citep{2018ApJ...859L..20D, 2023MNRAS.523.4136B}.
Its property depends sensitively on the inclination angle, i.e., it is faster and more dilute at a lower inclination angle (face-on with respect to the disk plane).

Nevertheless, the solid angle of the cloud is so small ($\sim$0.1) that the angular parts of the outflow that interact with it can be safely assumed to be isotropic.
Besides, our primary aim is to demonstrate the potential of our model, i.e., to check whether this model can explain the observations with reasonable parameter values.
So we do not take this inclination angle effect into account and consider the simplest case instead.
We take  the outflow to be conical and isotropic, with a half opening angle of $\theta_w=\pi/4$ (a solid angle of $\Omega_{w}\sim2$) \citep{2019MNRAS.483..565C}. The launching duration of  the outflow is $t_w\sim$ months.

The outflow is set to have a mean velocity of $v_w$, and the velocities of its front and rear ends are $v_{w}\pm av_{w}$,  respectively.
Then its radial width  grows as  $\Delta(r)\approx 2ar+\Delta_0$, where $\Delta_0= v_{w}t_w$ and $r=v_w t$.
Numerical simulations of accretion-driven outflows suggest a radial velocity spread of $a\approx0.1\sim0.6$ on particular inclination angles (see Figure 3 in \citeauthor{2018ApJ...859L..20D} \citeyear{2018ApJ...859L..20D}, or Figure 6 in  \citeauthor{2023MNRAS.523.4136B} \citeyear{2023MNRAS.523.4136B}).
Hereafter we set $a$=0.1.
The effect of a larger $a$ will be discussed at the end of  Section 2.4.

We define $\dot{m}(t^{\prime},r)$ as the outflow's density at the  elapsed time $t^{\prime} = t-R_{in}/v_w$ at distance $r$.
It is assumed to firstly increase linearly, then peak at $t_w(r)\equiv t_w\Delta(r)/\Delta_0$, and then decay in a power law of index $-\frac{5}{3}$:
\begin{equation} \label{eqn:L_k}
	\dot{m}_{w} (t^{\prime},r) = \frac{\dot{m}_0 \Delta_0}{\Delta(r)}\times\left\{
	\begin{aligned}
		& \frac{t^{\prime}}{t_{w}(r)},    & t^{\prime}<t_{w}(r),\\
		& \left[\frac{t^{\prime}}{t_{w}(r)}\right]^{-5/3},  & t_{w}(r)\le t^{\prime},
	\end{aligned}
	\right.
\end{equation}
where  $\dot{m}_{0}$ is the peak mass loss rate. The total mass of the outflow is $m_{w} = \int\dot{m}_{w}(t^{\prime},r) dt^{\prime}$.

\subsection{Radiation properties}

Assuming a fraction $\epsilon_{b}$ of the pre-shock outflow kinetic energy is   converted into the magnetic energy density $B^{2}/(8\pi) = \epsilon_{b}\rho_{w}v_{w}^{2}$ \citep{1998ApJ...497L..17S,2019MNRAS.487.4083Y}, the magnetic field strength is:
\begin{align}\label{eqn:B}
B & = \sqrt{8\pi\epsilon_{b}\rho_{w}} v_{w}\notag \\
& \approx 0.1\  \Omega_{w}^{-\frac{1}{2}}  \epsilon_{b,-3}^{\frac{1}{2}} \dot{m}^{\frac{1}{2}} _{w,-1}(t^{\prime},r)
\beta_{w,-1}^{\frac{1}{2}} R_{in,-1}^{-1}   \ [\mathrm{G}]
\end{align}
where $\epsilon_{b,-3} \equiv \epsilon_{b}/0.001$,
$\dot{m}_{w,-1}(t^{\prime},r) \equiv \dot{m}_{w}(t^{\prime},r)/(0.1\ \mathrm{M_{\odot}/yr})$,
and $R_{in,-1} \equiv R_{in}/0.1\ \mathrm{pc}$.
Here $\epsilon_{b}$  has a wide range from $10^{-4}$ to $10^{-1}$ \citep{2014PASA...31....8G,2014ApJ...785...29S,2015ApJ...806...15Z,2016MNRAS.461...51B}.

The peak synchrotron specific radiating  power (per unit frequency) from an electron with a Lorentz factor of $\gamma_e$ can be estimated as  $P_{\nu,max}=P(\gamma_{e})/\nu(\gamma_{e})= m_{e}c^{2}\sigma_{t}B/(3q_{e})$ and is independent of $\gamma_e$ \citep{1998ApJ...497L..17S}. Thus, the  peak specific flux of the swept-up electrons is $F_{\nu,max} = N_{e}P_{\nu,max}/(4\pi D^{2})$, where $D$ is the distance of the source and $N_e$ is the total number of emitting electrons.

\subsubsection{Time dependence of Radiating electrons}

The number of emitting electrons at any given time $t^{\prime}$ is:
\begin{align}
N_{e}(t^{\prime}) = \frac{\Omega_{c}}{\Omega_{w}}
\frac{\int_{0}^{t^{\prime}}  \dot{m}_{w}(t,r) dt}{m_p}   ,
\label{eqn:N_e}
\end{align}
where $\Omega_c$ is the cloud's solid angle and $\Omega_{c}/\Omega_{w}<1$ is the fraction of the outflow that indeed interacts with the cloud.

However, these electrons would cool due to both radiation and expansion (e.g., \citeauthor{1999ApJ...520..634D} \citeyear{1999ApJ...520..634D}).
The synchrotron cooling time for electrons with $\gamma_{e}$ is \citep{1998ApJ...497L..17S}
\begin{align}
t_{syn}& = \frac{6\pi m_{e}c}{\sigma_{t} \gamma_{e}B^{2}}\notag\\
& = 7.7\times 10^{10}\   \gamma_{e}^{-1} B^{-2}_{-1}\  [\mathrm{s}].
\label{eqn:t_syn}
\end{align}

The adiabatic cooling timescale $t_{ad}$ is generally estimated as the time required for the flow to cross the characteristic spatial scale  $ t_{dyn}\equiv R_{c}/v_{w} \approx 10^{7}R_{c,-2}\beta_{w,-1}$  s  where $R_{c,-2}\equiv R_{c}/0.01$ pc \citep{2019MNRAS.484.4760B}. 
However, the simulation of the outflow-cloud interaction  by \cite{2021MNRAS.507.1684M} suggests that in the case of a long-lasting outflow $t_w\gg t_{dyn}$, the  post-shock outflow stream is confined  rather than expanding freely, while it approaches the free expansion if $t_{w}\leq t_{dyn}$.
Therefore, we use the following fitting formula from \citeauthor{2021MNRAS.507.1684M} (\citeyear{2021MNRAS.507.1684M}, their Figure B2)  to estimate
\begin{equation} \label{eqn:t_ad}
	t_{ad}  = \left\{
	\begin{aligned}
		& t_{dyn},  &(t_{w}/t_{dyn})\le1,\\
		& 1.36 t_w - 0.36 t_{dyn},  & 1<(t_{w}/t_{dyn})<15,\\
		& 20t_{dyn},     &  15\le (t_{w}/t_{dyn}).
	\end{aligned}
	\right.
\end{equation}

The Lorentz factor of electrons which radiate mainly in GHz is about $\gamma_{e}\sim50$ for $B\sim$0.1 G, Then we have $t_{syn}\approx 10^{9} \ \mathrm{s}\gg t_{ad}$.
Thus,  the electron cooling is dominated by the adiabatic expansion.

Taking this cooling timescale into account, Eq. (\ref{eqn:N_e}) would be corrected as:
\begin{align}
N_{e}(t^{\prime}) = \frac{\Omega_{c}}{\Omega_{w}}
\frac{\int_{t^{\prime}-t_{ad}}^{t^{\prime}}  \dot{m}_{w}(t,r) dt}{m_p}   ,
\label{eqn:N_e_2}
\end{align}
for $t^{\prime}>t_{ad}$ since the electrons accelerated at a time of $t_{ad}$ earlier would have all cooled down.

\subsubsection{Synchrotron spectrum}

The synchrotron flux density at any given frequency can be calculated with $F_{\nu,max}$ and the spectral shape \citep{1998ApJ...497L..17S}.
Electrons are usually accelerated to a power-law distribution as $\mathrm{d}N(\gamma_{e}) = C \gamma_{e}^{-p} \mathrm{d}\gamma_{e}, \gamma_{e}>\gamma_m$ where $p=2.5$ is the spectral index and $\gamma_{m}$ is the minimum Lorentz factor of the electrons \citep{1997ApJ...490..619S,1998ApJ...492..219G,1998ApJ...497L..17S}. 
The coefficient $C=(p-1)\gamma_m^{p-1}N_e(t^{\prime})$ is obtained with $\int_{\gamma_m}^{\infty}dN(\gamma_e)=N_e(t^{\prime})$.
Assuming a fraction $\epsilon_{e}$ of the outflow's kinetic energy goes into accelerating electrons,
$\gamma_m$ can be calculated, based on the total energy of the accelerated electrons, as: $\gamma_{m} = \frac{1}{2}\epsilon_{e} (m_{p}/m_{e}) [(p-2)/(p-1)] \beta^{2}_{w}$.

The spectrum produced by these power-law electrons is featured with three breaking frequencies.
The first breaking frequency is the  characteristic frequency $\nu_{m} = \gamma_{m}^{2}eB/2\pi m_{e}c
  \approx 1\times 10^{6} B_{-1}\gamma_{m,2}^{2} $ Hz
where $\gamma_{m,2} \equiv \gamma_{m}/2$.
The second is the self-absorption frequency $\nu_{a}$, below which the emitting region is optically thick.
It can be obtained by solving $\int \alpha_{\nu} \mathrm{d}l\approx \alpha_{\nu}l= 1$ where
$l$ is the radial width of the emitting region and $\alpha_{\nu} \sim 10^{13} (C/\Omega_c R_{in}^{2} l) B^{(p+2)/2} \nu^{-(p+4)/2}$ is the self-absorption coefficient \citep{1986rpa..book.....R}.
Then we have:
\begin{align}
\nu_{a}  \approx 5.4\times 10^{9}\  &\Omega_{w}^{-\frac{p+6}{2(p+4)}}   
\epsilon_{e,-1}^{\frac{2}{p+4}}
\epsilon_{b,-3}^{\frac{p+2}{2(p+4)}}
 \gamma_{m,2}^{\frac{2(p-2)}{p+4}}
 \notag\\
 &\dot{m}_{w,-1}^{\frac{p+6}{2(p+4)}}
  \beta_{w,-1}^{\frac{p+10}{2(p+4)}} 
 R_{in,-1}^{-\frac{p+6}{p+4}}   
 t_{2}^{\prime \frac{2}{p+4}}
  \ [\mathrm{Hz}].
\end{align}
where $t^{\prime}_2 \equiv t^{\prime}/100$ days.
The third is the cooling frequency  $\nu_{c}$, where electrons with Lorentz factor $\gamma_{c}$ would have cooled down in the time $t^{\prime}$.
It can be obtained by solving $\gamma_{c}m_{e}c^{2}=P(\gamma_{c})t^{\prime}$ \citep{1998ApJ...497L..17S}, which gives $\nu_{c} = 1.6\times 10^{13}B^{-3}_{-1} t^{\prime -2}_2$ Hz.

\begin{figure}[htp]
\includegraphics[width=0.48\textwidth]{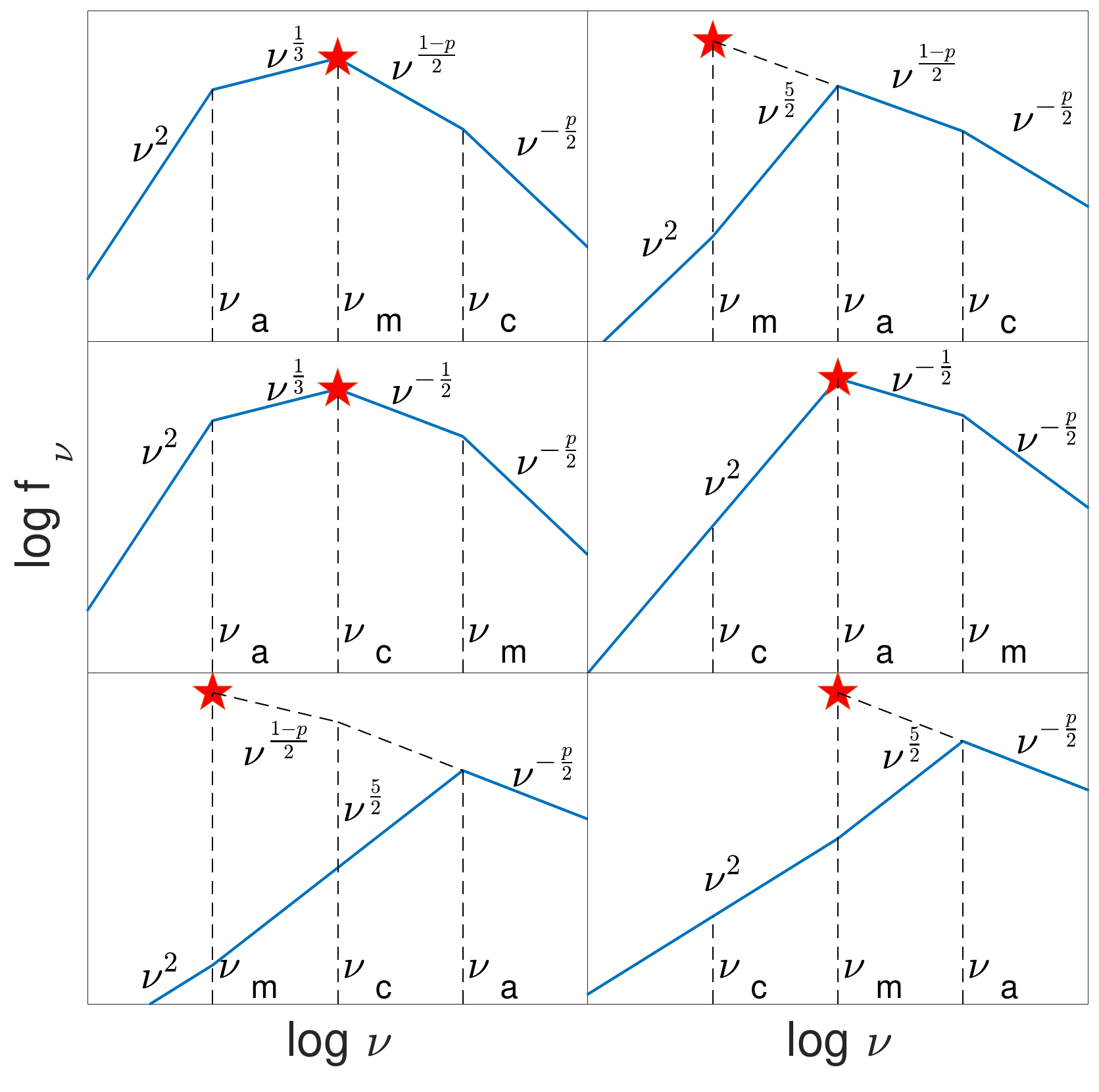}
\caption{The possible synchrotron spectra from the shocked electrons.
The spectral shape is determined by the relative ordering of the self-absorption frequency $\nu_{a}$, the characteristic frequency $\nu_{m}$, and the cooling frequency $\nu_{c}$.
The red star marks $F_{\nu,max}$.
\label{fig:spectra}}
\end{figure}

Figure \ref{fig:spectra} shows all the six possible spectra cases, each of which consists of three or four power law segments, and the red star marks $F_{\nu,max}$ \citep{2002ApJ...568..820G,2013MNRAS.435.2520G}. 
Comparison of these frequencies shows that the upper right case ($\nu_m<\nu_a<\nu_c$) in Figure \ref{fig:spectra} is concerned most of the time.
Finally, we smooth our analytical spectra with the following equation \citep{2002ApJ...568..820G}:
\begin{align}
F_{\nu} &= F_{\nu}(\nu_m) \left[ (\frac{\nu}{\nu_m})^{2} exp(-s_{4}(\nu/\nu_{m})^{2/3}) + (\frac{\nu}{\nu_m})^{5/2}   \right] \notag\\
& \times  \left[  1+ (\frac{\nu}{\nu_a})^{s_5 (\beta_2-\beta_3)}   \right]^{-1/s_5},
\end{align}  
where $\beta_2 = 5/2$, $\beta_3 = (1-p)/2$, $s_4 = 3.63p-1.6$ and $s_5 = 1.25-0.18p$.

\subsection{Light curve from the shocked outflow}

The radio flare from the outflow-cloud interaction would appear at  $R_{in}/v_{w}\sim$0.1 pc/0.1 c $\sim$ year after TDEs. 
Its  light curve first rises as the outflow density upstream the bow shock increases with time. 
Then it peaks at $t_w(R_{in})$ when the outflow's density starts declining while $N_{e}(t^{\prime})$ remains nearly unchanged.
Later, the decrease in $N_{e}(t^{\prime})$ after $t_{ad}$ due to the adiabatic expansion would make $F_{\nu}$ decrease more rapidly.
The peak radio luminosity at time $t^{\prime}=t_{w}(R_{in})$ can be crudely estimated as:

\begin{align}
(\nu L_{\nu})_{peak} & \approx \nu_{a}N_{e}P_{\nu,max}  (\frac{\nu_a}{\nu_m})^ {\frac{1-p}{2}}\notag \\
& \approx 8\times10^{38} \  \Omega_{c}      \Omega_{w}^{-\frac{19+3p}{2(p+4)}}  
						          \epsilon_{e,-1}     \epsilon_{b,-3}^{\frac{3p+5}{2(p+4)}}    \gamma_{m,2}^{\frac{4p-11}{p+4}}    \notag\\
&   \dot{m}^{\frac{3p+19}{2(p+4)}}_{0,-1}  \beta_{w,-1}^{\frac{7p+21}{2(p+4)}}  R_{in,-1}^{-\frac{2p+22}{2(p+4)}}  t_{w,2}^{\prime \frac{7}{p+4}}(R_{in}) \ [\mathrm{erg/s}].
\label{eqn:L_peak}
\end{align}
where $t_{w,2}^{\prime}(R_{in})\equiv t_{w}(R_{in})/100$ days.
It shows that the luminosity depends most sensitively on $v_{w}$: a faster outflow would generally produce a brighter flare.

\begin{figure}
\centering
\includegraphics[width=0.45\textwidth]{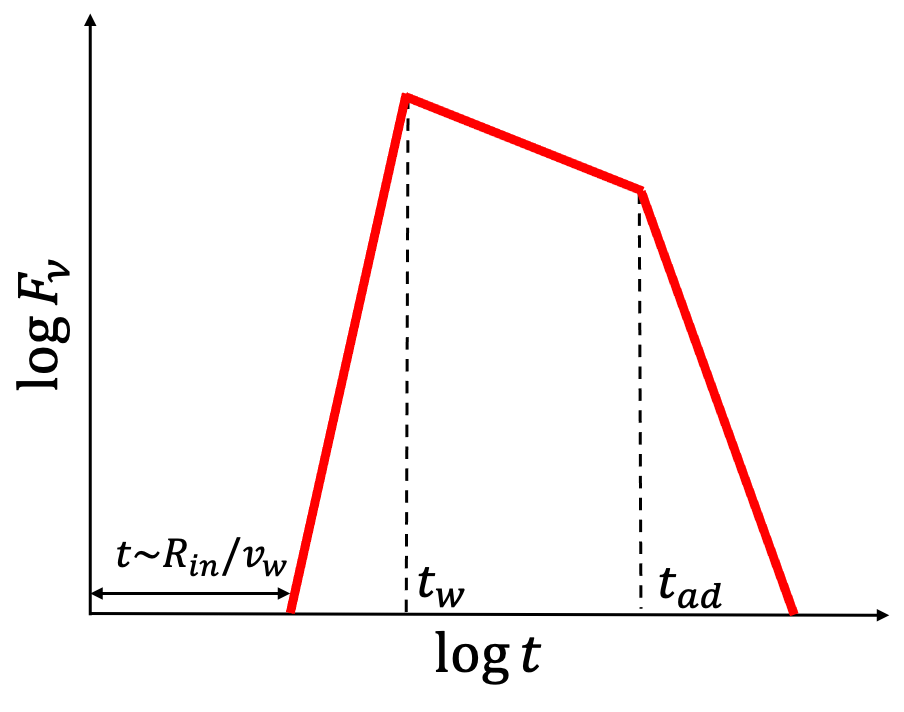}
\caption{Sketch of the predicted radio light curve from the outflow-cloud interaction where $t_w$ is the duration of the outflow and $t_{ad}$ is the adiabatic cooling timescale of the shocked outflow.
\label{fig:wind_clouds}}
\end{figure}

Figure \ref{fig:wind_clouds} shows the predicted light curve from the outflow-cloud interaction. The rising timescale of $t_w(R_{in})$ is determined by the width of the outflow $\Delta(R_{in})$.
A larger velocity spread of the outflow would cause a wider width, thus a shallower rise of the light curve.
Since $t_{w}(R_{in})$ is smaller than $R_{in}/v_w$, the outflow-cloud interaction can reproduce the sharp rise of the light curve due to the time lag of the collision.
If one were to use a power-law form $t^k$ to fit this rise, then the index $k$ can be crudely estimated as $k = log(F_{\nu,peak}/F_{\nu,ul})/log[1+t_w/(R_{in}/v_w)]\approx 1/log(1+\Delta_0/R_{in})$, where  $F_{\nu,ul}$ is the last pre-flare observational upper limit and is generally an order of magnitude lower than the flare peak flux density $F_{\nu,peak}$.
 Thus, we have $k\approx3\sim20$ when $\Delta_0/R_{in}$ is $0.1\sim1$.

\section{The outflow-CNM interaction}\label{CNM}

The sweeping-up of a diffuse CNM by the outflow would produce radio emission as well.
Furthermore, changes in the velocity of the outflow due to this process would affect the radiation properties from the bow shock.
Here we assume a CNM density profile  $n_{CNM}(r) = A r_{-2}^{-1}\  \mathrm{cm^{-3}}$ where $A$ is a constant and $r_{-2}\equiv r/10^{-2}$ pc, similar to the density distribution of CNM around Sgr $A^{\star}$ \citep{2006ApJ...640..319X,2019ApJ...871..126G}.
We model the hydrodynamic evolution of the outflow under a set of  hydrodynamical equations \citep{2000ApJ...543...90H} :
\begin{align}
\frac{dr}{dt} & = \beta_{w} c\\ 
\frac{dM}{dr}  & = \Omega_w r^{2}n_{CNM} m_{p},\\
\frac{d\beta_w}{dM} & = - \frac{ \beta_w(1+\beta^2_w/4) }{ m_{w}+ M+(1-\epsilon)(1+\beta^2) M},
\end{align}
where   $M$ is the swept CNM mass by the forward shock and $\epsilon$ describes the radiation efficiency \citep{1999ApJ...520..634D}. 

Integrating these equations with proper initial conditions, we can obtain the hydrodynamic evolution of the outflow, and calculate the associated radiation properties from the shocked CNM according to \cite{1998ApJ...497L..17S}.
The resultant radio luminosity would rise  as $\nu L_{\nu}\propto t$, and reach its peak luminosity of
\begin{align}
 (\nu L_{\nu})_{peak} &= 1.8\times 10^{36} m_{w,-1}^{\frac{p+11}{4(p+4)}}
 A_{4}^{\frac{5p+27}{4(p+4)}}
 \Omega_{w}^{\frac{3p+5}{4(p+4)}}\notag\\
 &
 \beta_{w,-1}^{\frac{17p-9}{p+4}}
\epsilon_{b,-3}^{\frac{3p+5}{2(p+4)}}
\epsilon_{e,-2}^{\frac{10p-7}{p+4}} \  [\mathrm{erg/s}] 
\label{eqn_Lcnm}
\end{align}
at the deceleration time $t_{dec}= 3.5\times 10^{3}  m_{w,-1}^{1/2}
 A_{4}^{-1/2}
 \Omega_{w}^{-1/2}
 \beta_{w,-1}^{-1} $ [day],
 when the mass of the shocked CNM equals to that of the outflow, and $A_{4}\equiv A/10^{4}$, $m_{w,-1}=m_{w}/0.1$ $\mathrm{M_{\odot}}$.

\begin{figure}[htp]
\centering
\includegraphics[width=0.45\textwidth]{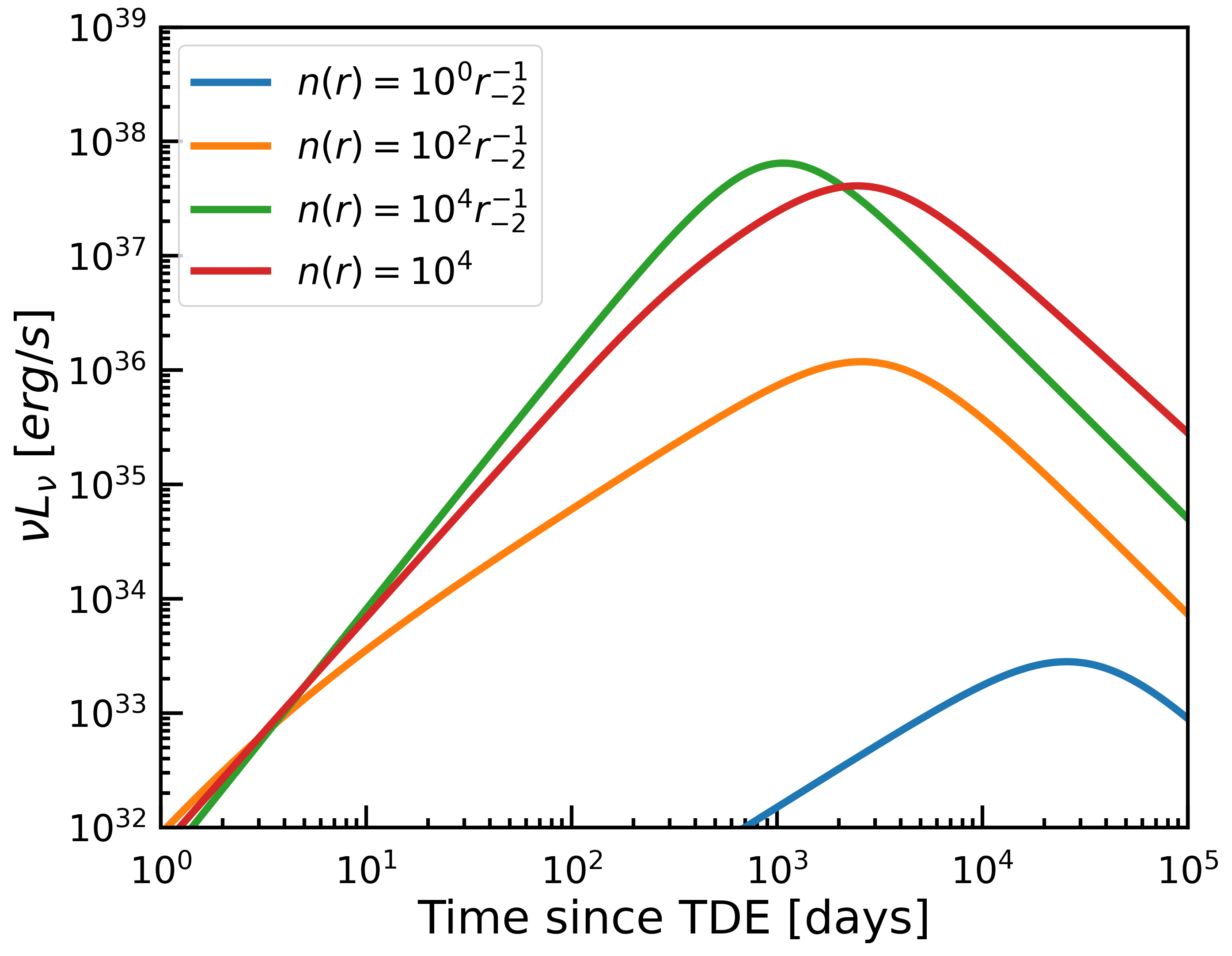}
\caption{Predicted light curves at 1 GHz from the outflow-CNM interaction for different CNM density profiles. Other parameters as $\epsilon_e=0.01$, $\epsilon_b=0.001$, $m_{w}=0.1\ \mathrm{M_{\odot}}$ and $v_{w}=0.3$ c.}
\label{fig:lcs}
\end{figure}

Figure \ref{fig:lcs} shows the predicted light curves at 1 GHz for different CNM density  profiles.
It  shows that a denser CNM (i.e., a higher A) produces a brighter and earlier radio flare, while a more dilute CNM leads to a dimmer, thus maybe undetectable, flare ($\nu L_{\nu}<10^{36}$ erg/s).

\section{Application}

\begin{table*}[!htbp]
\centering
\caption{Parameters for the outflow-cloud interaction model applied to the late radio flares in five TDEs, as well as those for the outflow-CNM interactions applied to the early radio flare. The multiple flares in ASASSN-15oi require five clouds (see Figure \ref{fig:15oi}) while it is two clouds for AT2019azh (see Figure \ref{fig:AT2019azh}). }
	\begin{tabular}{ccccccccccc}
	\toprule
	& CNM & \multicolumn{2}{c}{Forward shock} & \multicolumn{3}{c}{Outflow} &
	\multicolumn{2}{c}{Cloud} & \multicolumn{2}{c}{Bow shock}  
	\\
	 &$A$ &   $\epsilon_{e}$ &$\epsilon_{b}$   &$v_w$   & $m_w $ & $t_w$  &   $R_{in} $    & $R_{c}$& 	$\epsilon_e$ & $\epsilon_b$ \\
	&[$10^{3}$ $\mathrm{cm^{-3}}$] & [$10^{-2}$] &[$10^{-3}$] & [$c$]& $ [10^{-2}\ \mathrm{M_{\odot}}]$ &   [$10^{2}$\ day]  & [$10^{-1}$\ pc]  &[$10^{-1}$\ pc]&$[10^{-2}]$ &$[10^{-3}]$ \\
	\midrule
	AT 2020vwl           &5.8  &  2   &1&   0.24 & 0.6  &0.4 &   1.1  &0.61 &20&200    \\ 
	AT 2019azh 	      &8.5  & 1    &1     & 0.46   & 0.45&1.64  & 1.06  & 0.38 &  5& 40 \\
	AT 2019azh(2)	      &  &     &    &    &  &   & 2 & 0.8 &  20& 140 \\
	IGR J12580+0134   &7  &  1   &1     & 0.41  & 4.4 & 0.8 &  4.23 & 1.9 &  12 & 120 \\
	AT 2018hyz           &    0.05    &1&1& 0.6  &31 & 5.2 &  4.2  & 2.6 &  5 & 1\\
	ASASSN-15oi(1)     &   &&&     0.5  & 3.5  & 0.23 &  0.45 & 0.13  &  1& 1.5 \\
	ASASSN-15oi(2)    &      &&&   0.47 &         &  &  0.65  & 0.33   &  0.5 & 1 \\
	ASASSN-15oi(3)    &     &&&   0.44 &        & &  1& 0.37   &  3 & 1 \\
	ASASSN-15oi(4)    &     &&&     0.4&       &  & 3.8& 1.7 &  10 & 50 \\
	ASASSN-15oi(5)    &      &&&   0.36&        &   & 6.4 &  2.5& 13  &  100\\	
	\bottomrule
	\end{tabular}
\label{tab:pam}
\end{table*}

\begin{figure}
\centering
\includegraphics[width=0.5\textwidth]{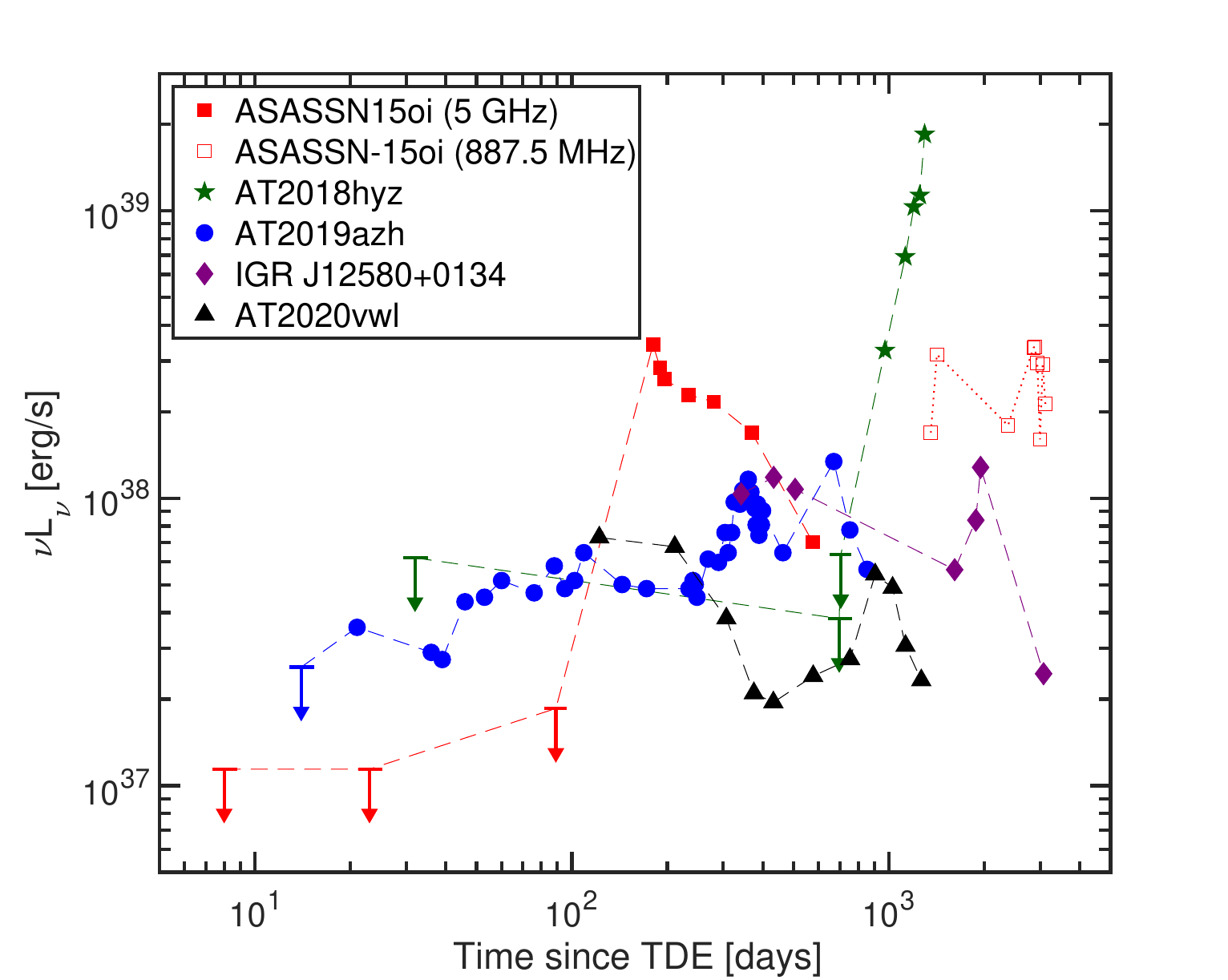}
\caption{Light curves of TDEs with late-time radio flares: ASASSN-15oi (5 GHz, \citeauthor{2021NatAs...5..491H} \citeyear{2021NatAs...5..491H}; 887.5 MHz \citeauthor{2024ATel16502....1A} \citeyear{2024ATel16502....1A};  the data at 1,414 and 2,376 days are obtained with $f_{887.5\  \mathrm{MHz}} = f_{3\  \mathrm{GHz}} \left[887.5\  \mathrm{MHz}/3\  \mathrm{GHz}\right]^{1/3}$ for they are observed by VLASS at about 3 GHz.), AT2018hyz (5 GHz, \citeauthor{2022ApJ...938...28C} \citeyear{2022ApJ...938...28C}), AT2019azh (15.5 GHz, \citeauthor{2022ApJ...933..176S} \citeyear{2022ApJ...933..176S};  \citeauthor{2022MNRAS.511.5328G} \citeyear{2022MNRAS.511.5328G}), IGR J12580+0134 (1.5 GHz, \citeauthor{2022ApJ...925..143P} \citeyear{2022ApJ...925..143P}) and AT2020vwl (5.5 GHz, \citeauthor{2023MNRAS.tmp.1199G} \citeyear{2023MNRAS.tmp.1199G}, \citeauthor{2023ATel16165....1G} \citeyear{2023ATel16165....1G},\citeauthor{2024arXiv241018665G} \citeyear{2024arXiv241018665G} ), including early upper limits.
\label{fig:lc_sample}}
\end{figure}

In this section, we apply our model to five TDE candidates with late-time radio flares: AT2018hyz, AT2019azh, ASASSN-15oi, IGR J12580+0134 and AT2020vwl.
Their radio light curves are shown in Figure \ref{fig:lc_sample}, with peak luminosities of $10^{37}\sim 10^{39}$ erg/s.

Besides showing late-time  radio flares, two events, AT2019azh and AT2020vwl, show an earlier peak (t$<$200 days) as well.
A third event, IGR J12580+0134, shows a slow decline before the late, rapid rebrightening.
Noticing their apparent slow evolution, we would interpret these early flares or decline as being due to the outflow-CNM interaction.
Note that the early radio flare would be missing if the CNM is dilute, which we think to be the case for the other events.

\subsection{AT2020vwl}

\begin{figure}[ht]
\centering
\includegraphics[width=0.45\textwidth]{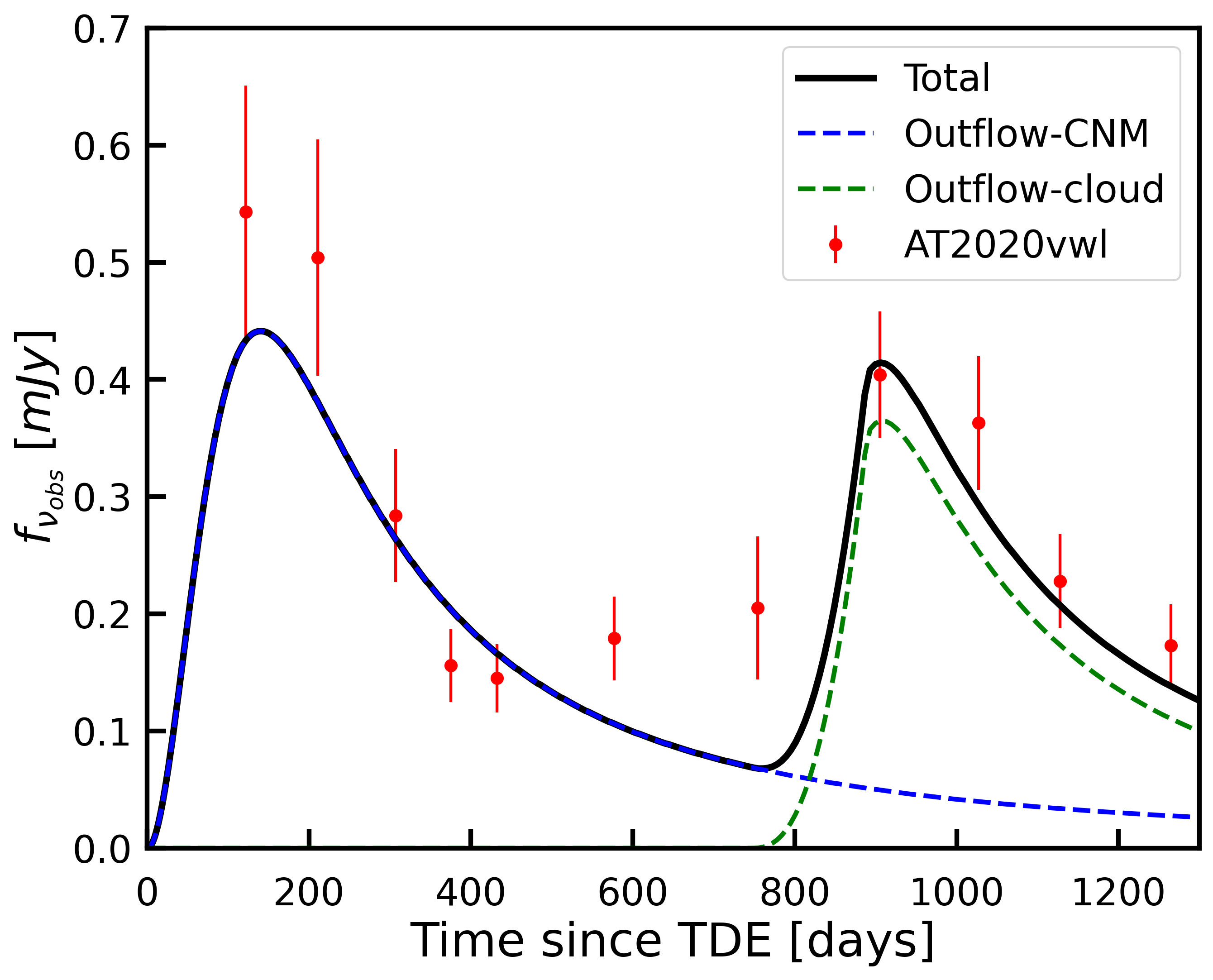}
\caption{The observed and modeled Light curves at 5.5 GHz for AT2020vwl. Here we consider the emissions from the outflow-CNM interaction (blue dashed line), and that from the outflow-cloud interaction  (green dashed line), respectively.
The combined flux is shown as the black solid line.
\label{fig:AT2020vwl}}
\end{figure}

\begin{figure}[ht]
\centering
\includegraphics[width=0.405\textwidth]{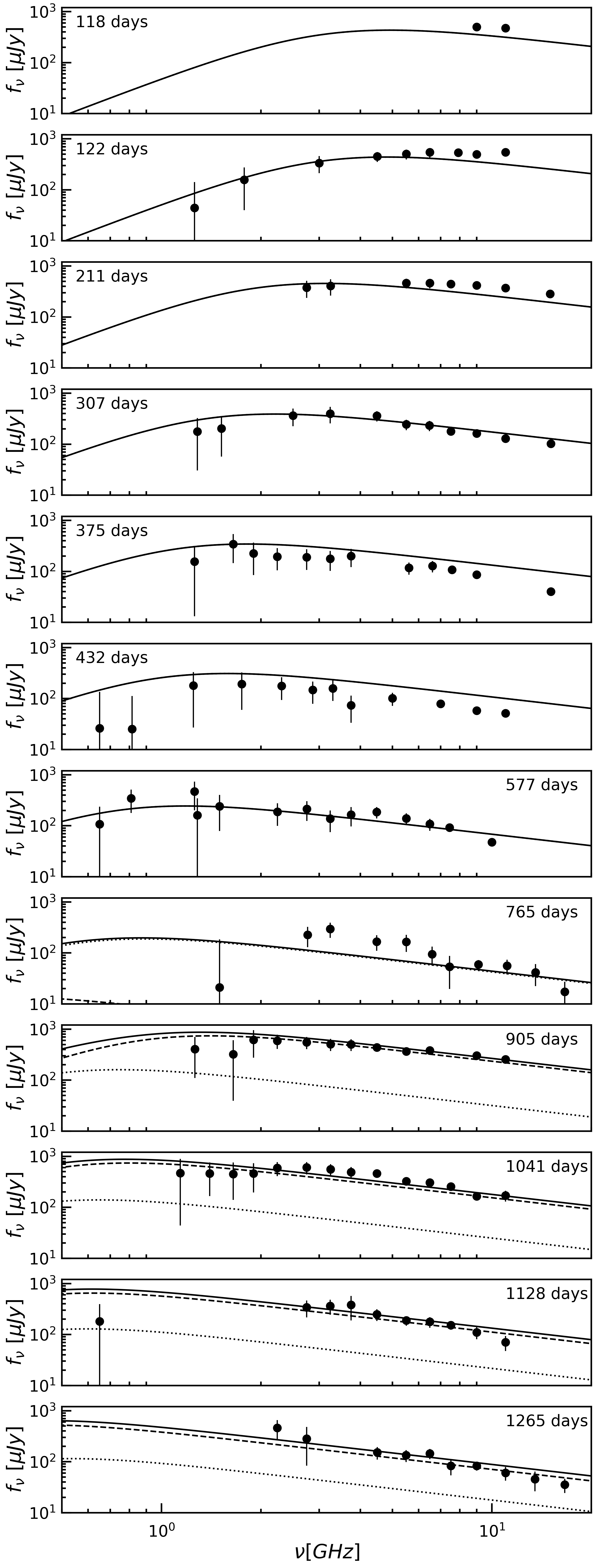}
\caption{Model fittings of the spectra and its evolution for AT2020vwl. The radio data is from \cite{2023MNRAS.tmp.1199G} and \cite{2024arXiv241018665G}.
The early emission ($t<800$ days) is dominated by the outflow-CNM interaction (dotted line) and the later one by the outflow-cloud  interaction (dashed line). The solid line is the sum of the two emission components.
\label{fig:AT2020vwl_s}}
\end{figure}

Figure \ref{fig:AT2020vwl} shows the observed and modeled light curve at 5 GHz for AT2020vwl.
The adopted parameters are listed in Table \ref{tab:pam}.
The green dashed line shows that the outflow-cloud interaction can generate a bright late-time radio flare after 600 days.
Figure \ref{fig:AT2020vwl_s} shows the model fittings of spectra by the outflow-CNM and the outflow-cloud  interaction.
The dotted line is from the outflow-CNM interaction while the dashed line is from the outflow-cloud interaction.
The solid line is the sum of them and the spectra are well explained by our model.

\subsection{AT2019azh}
Similar to AT2020vwl, the 15.5 GHz radio observation in AT2019azh shows an early rise of about 30 days and is followed by a late-time brighter radio flare with a peak luminosity of $10^{38}$ erg/s at about 350 days \citep{2022ApJ...933..176S}.
Even later (t$>$400 days) observation at $1\sim10$ GHz by \cite{2022MNRAS.511.5328G} reveals that the light curve at 15.5 GHz would have re-brightened again at $t\sim600$ days.
These would suggest that the outflow might have collided with two clouds, one inner and the other outer in distance.

\begin{figure}
\centering
\includegraphics[width=0.45\textwidth]{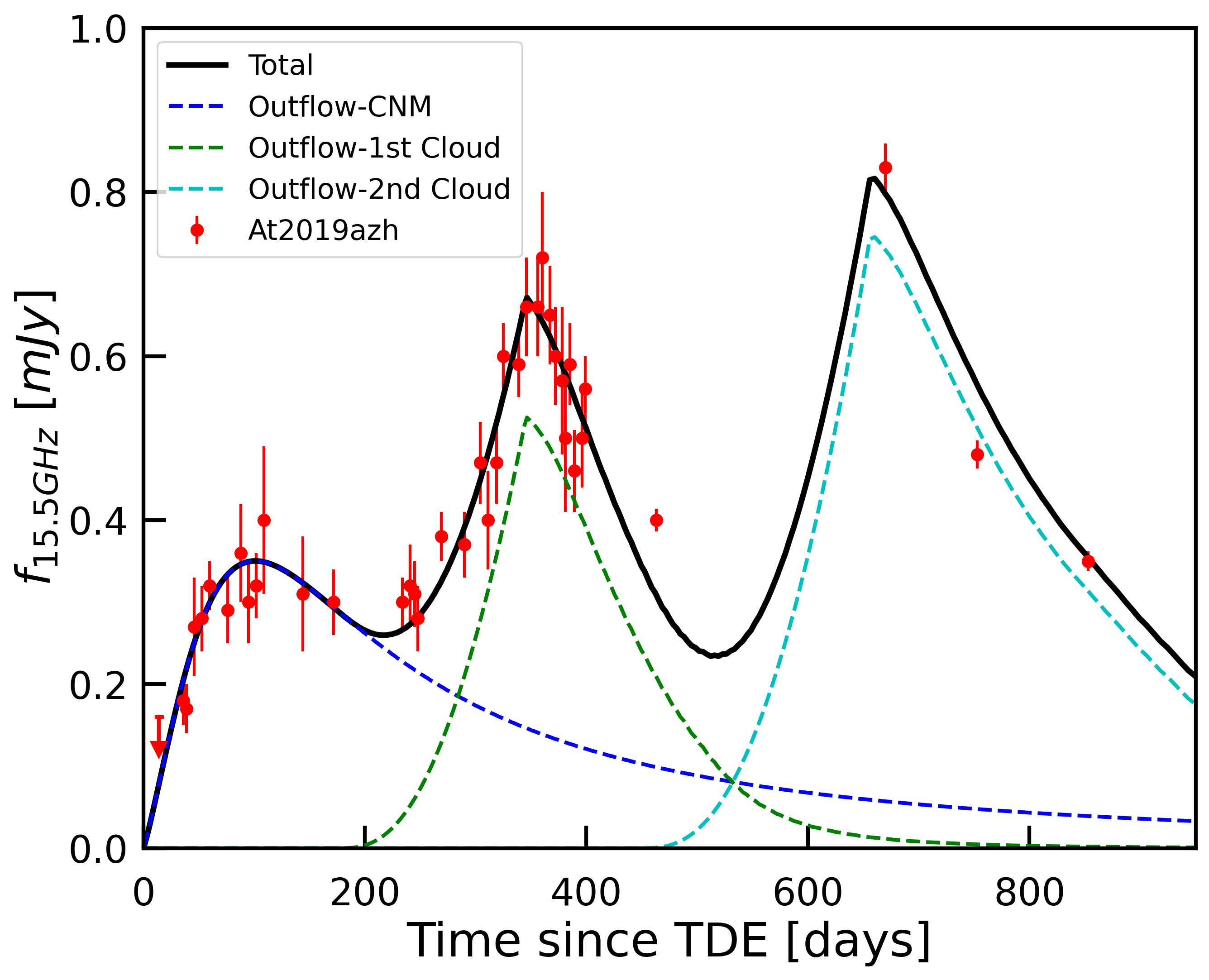}
\caption{Same as Figure \ref{fig:AT2020vwl} but for AT2019azh at 15.5 GHz,  including early upper limit. Here, two clouds are needed, for the second and the third flares, respectively.
\label{fig:AT2019azh}}
\end{figure}

Figure \ref{fig:AT2019azh} shows the observed and modeled light curves at 15.5 GHz for AT2019azh. 
The later (t$>$400 days) observational data has been extrapolated from $1\sim10$ GHz to 15.5 GHz using a power-law spectral model.
The overall model parameters adopted are listed in Table \ref{tab:pam}.

The outflow-CNM interaction produces a radio flare that peaks at about 100 days.
Though the outflow is decelerating,  its velocity is higher than 0.2 c while colliding with the clouds.
So when the clouds are hit, with higher  values (bath $\sim$ 0.1) of $\epsilon_b$ and $\epsilon_b$, the late-time and brighter radio flares are generated.
The power-law index of the temporal decay of the outflow mass rate $\dot{m}(t^{\prime},r)$ determines how fast the flare decays and we found an index of 2 is preferred over $5/3$.

\subsection{IGR J12580+0134}

\begin{figure}
\centering
\includegraphics[width=0.45\textwidth]{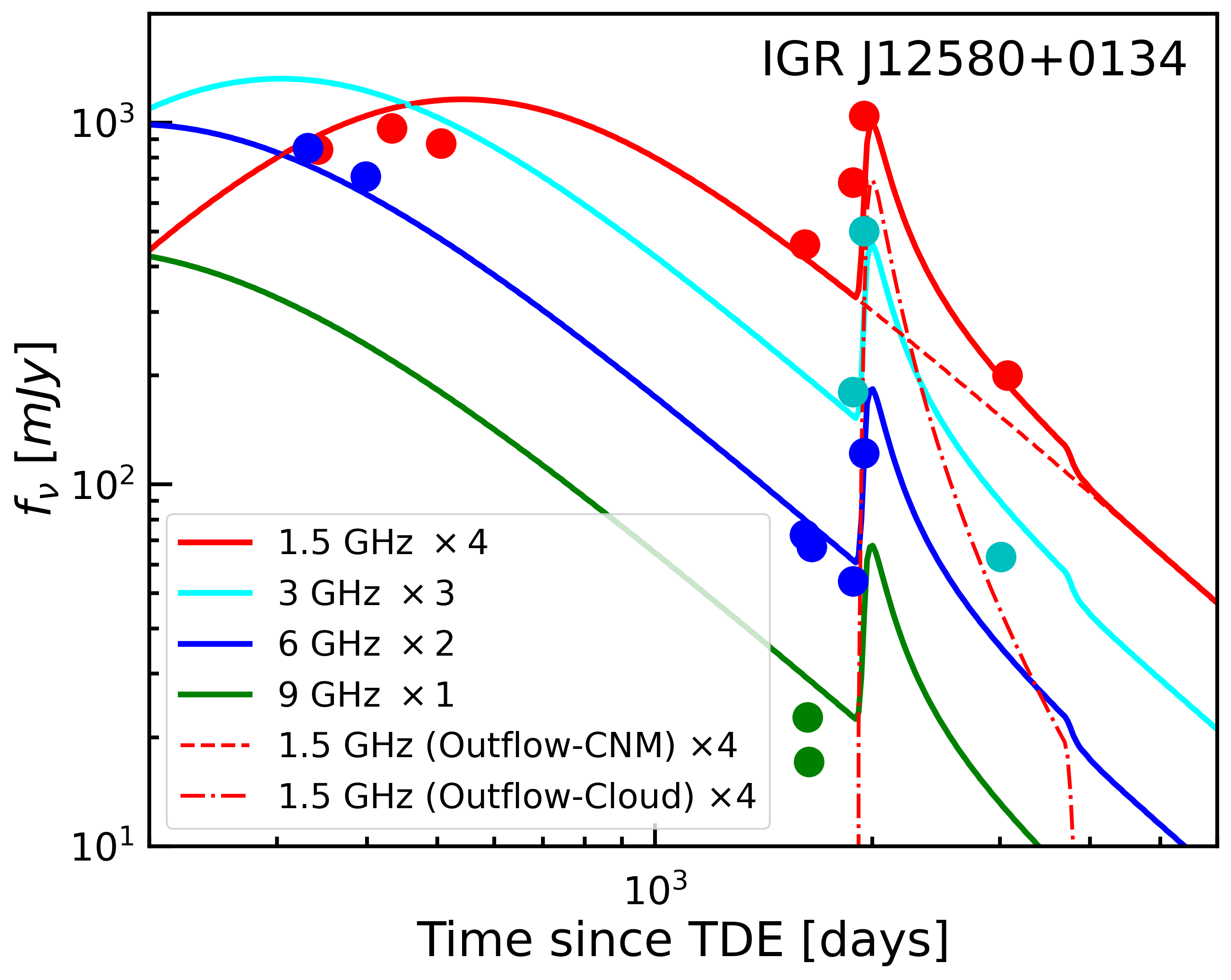}
\caption{Same as Figure \ref{fig:AT2020vwl} but for IGRJ12580+0134 at different frequencies.
\label{fig:IGRJ}}
\end{figure}

The radio emission of IGR J12580+0134 was detected 350 days after the X-ray outburst and showed a decline, while later observation at 1,300 days revealed a second flare in the radio with a peak luminosity of $10^{38}$ erg/s \citep{2022ApJ...925..143P}.
Figure \ref{fig:IGRJ} shows the comparison of our model with the observed light curves at different frequencies (all are shifted by a factor  for  clarity). The parameters adopted are listed in Table \ref{tab:pam}.
Here to accommodate the very steep rise of the late flare in this event, we set a=0 so as to limit the outflow's radial width.

For the outflow-CNM interaction as the origin of the early decline, the light curves for $
\nu<\nu_a$ would peak  when $\nu_a$ crosses $\nu$.
As for the outflow-cloud interaction that explains the late flare, the predicted light curves are expected to peak at  $t^{\prime} =t_{w}(R_{in})$ because $\nu>\nu_a$ is always the case.

\subsection{AT2018hyz}

\begin{figure}[htp]
\centering
\includegraphics[width=0.45\textwidth]{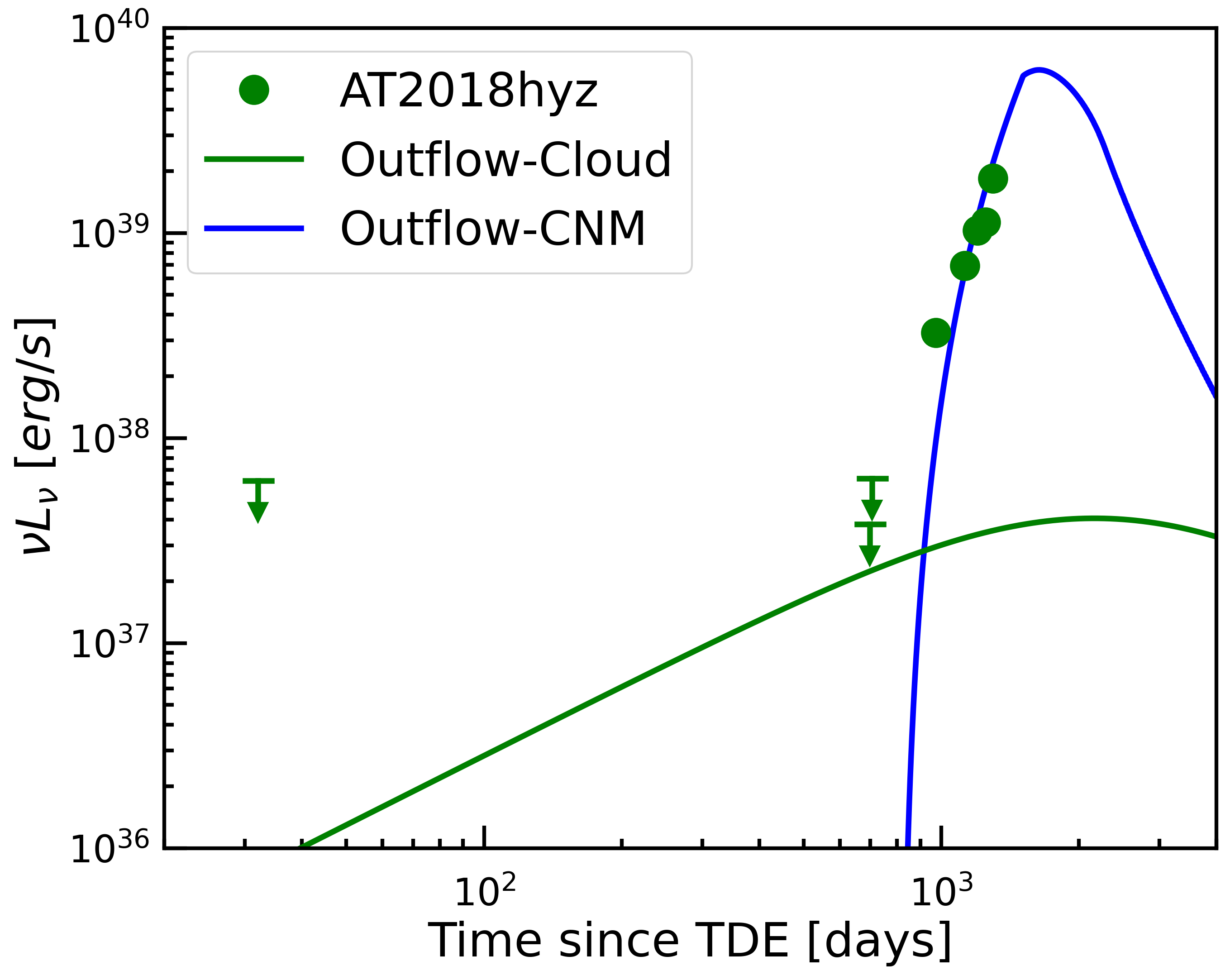}
\caption{The observed and modeled light curves of AT2018hyz at 5 GHz. The green line (setting $A=50$) is  one of the possible light curves from the outflow-CNM interaction that are consistent with the early non-detection upper limits.
\label{fig:AT2018hyz_5}}
\end{figure}

\begin{figure}[htp]
\centering
\includegraphics[width=0.45\textwidth]{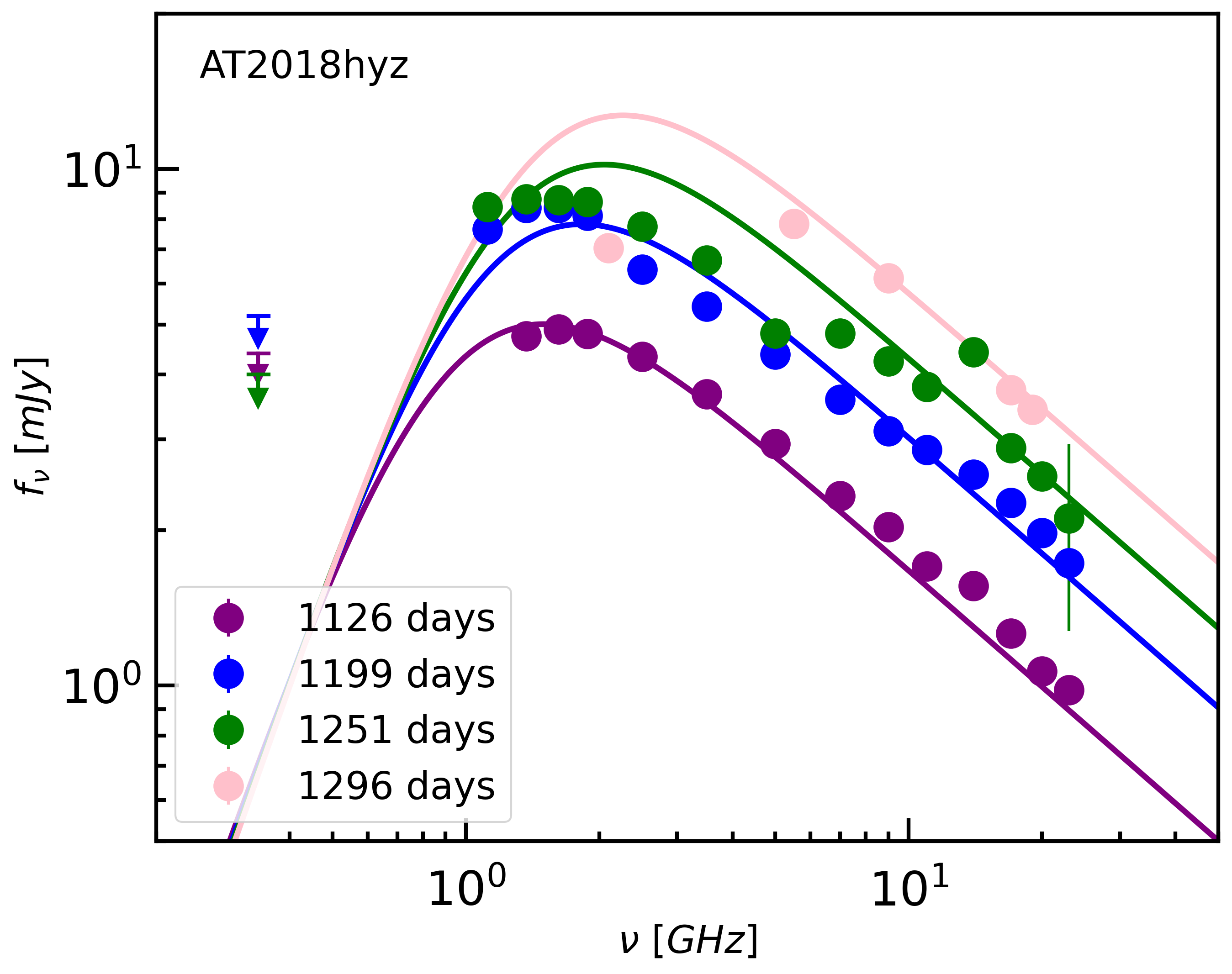}
\caption{Model fittings of the temporal evolution of the spectra from the outflow-cloud interaction for AT2018hyz,  including upper limits.
\label{fig:AT2018hyz}}
\end{figure}

The late-time radio flare in AT2018hyz appeared almost 1000 days after the optical discovery with a luminosity of $10^{39}$ erg/s \citep{2020MNRAS.497.1925G,2022ApJ...938...28C}.
Figure \ref{fig:AT2018hyz_5} shows the observed radio light curve
 and that predicted by our outflow-cloud interaction model, while Figure \ref{fig:AT2018hyz} shows the modeling of the spectra. 
The steep rise of the flare is well reproduced.
The model also shows a rapid decay after the peak of the flare, if there is no additional outflow-cloud interaction. 

The early radio non-detection of AT2018hyz, with a luminosity upper limit of $\nu L_{\nu}< 6\times 10^{37}$ erg/s at $t = 700$ days, hints at a very dilute CNM (see Section \ref{CNM}). 
Simply requiring that $(\nu L_{\nu})_{peak}$ of the outflow-CNM interaction be below this upper limit, that would mean $A < 430$ according to  Eq. (\ref{eqn_Lcnm}), with all other  parameters  adopting values given from fitting the late radio flare.
Note that the observed frequency in AT2018hyz may not be equal to the spectral peak frequency (usually $\nu_a$), so this constraint on $A$  might be very crude. 
Alternatively,  we can numerically  calculate the light curve from the shocked CNM and compare it with the early non-detections. 
We tried various values of $A$ and found that $A = 50 $ can give a result consistent with the data, which is shown in the green line in Figure \ref{fig:AT2018hyz_5}.
Therefore, $A<  50$ is required for AT2018hyz.

\subsection{ASASSN-15oi}

\begin{figure}
\centering
\includegraphics[width=0.45\textwidth]{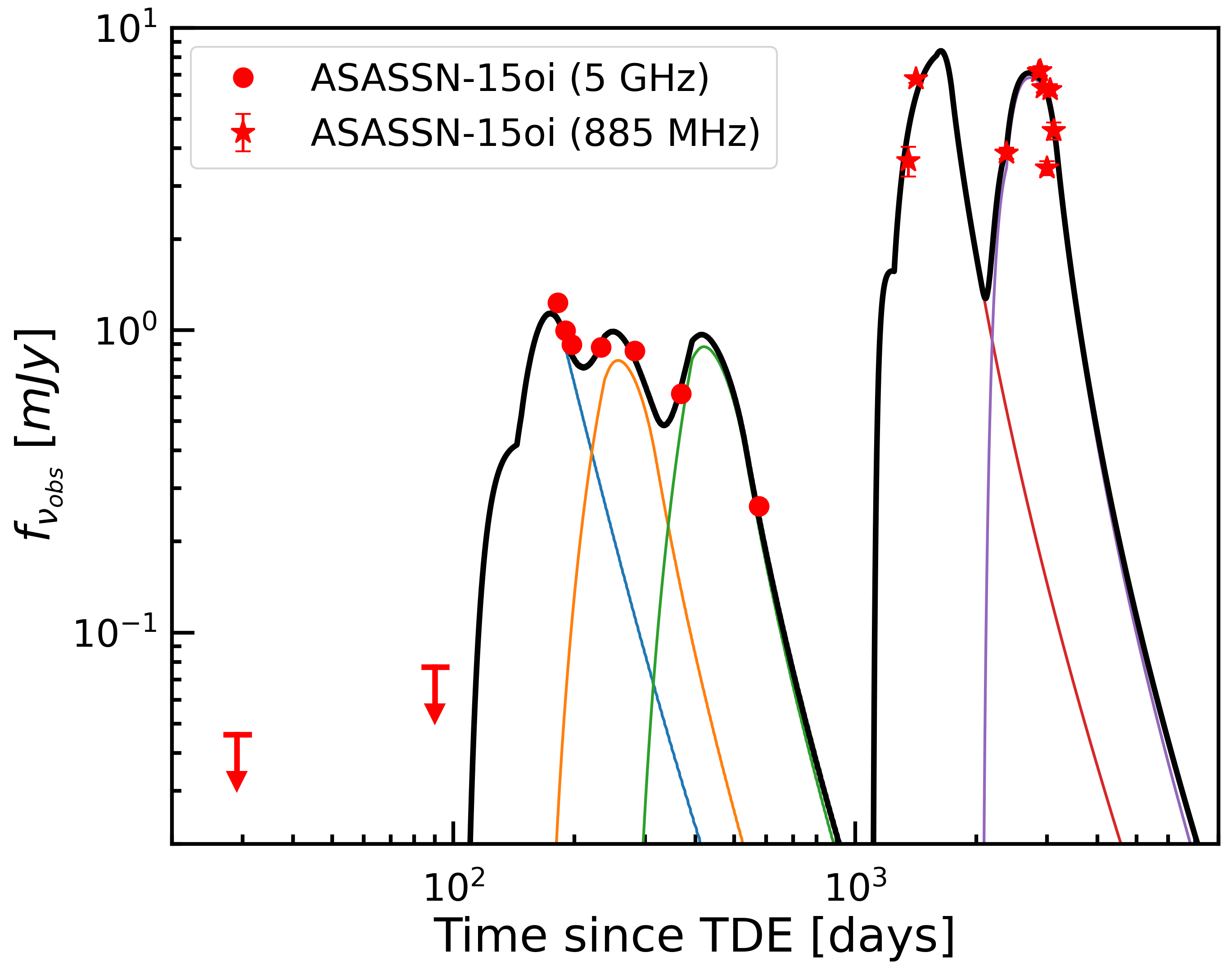}
\caption{The modeling of radio light curves at 5 GHz and 885 MHz for ASASSN-15oi,  including early upper limits.
The emission is from the interaction of the outflow with five giant clouds at different distances.
The different color thin lines mark the light curve from different clouds while the thick black line marks the total light curve.
The parameters adopted are listed in Table \ref{tab:pam}.
\label{fig:15oi}}
\end{figure}

\begin{figure}
\centering
\includegraphics[width=0.47\textwidth]{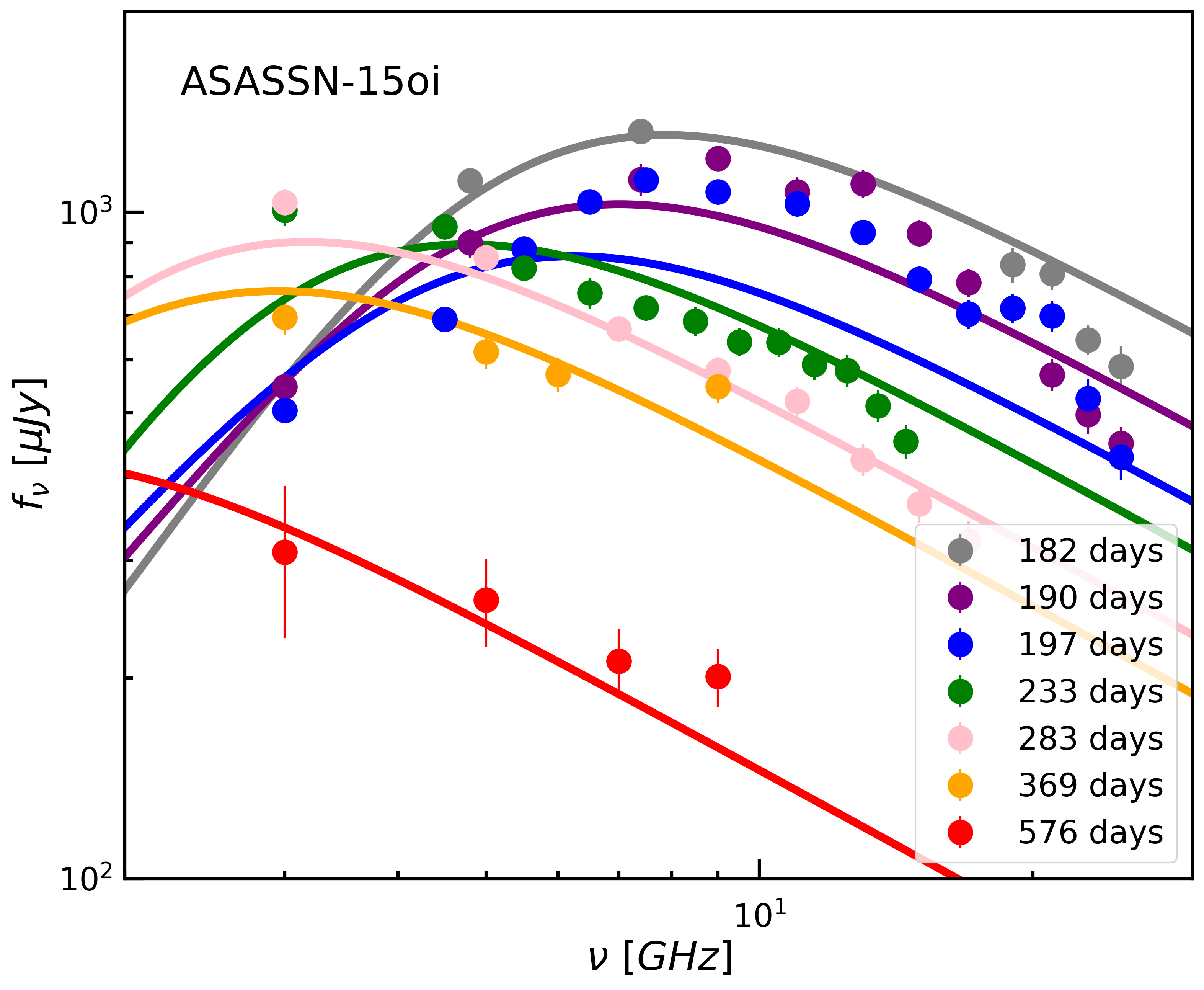}
\caption{Model fittings of the temporal evolution of the spectra from the outflow-clouds interaction for ASASSN-15oi. 
\label{fig:15oi_s}}
\end{figure}

Multiple radio flares are  observed in ASASSN-15oi \citep{2021NatAs...5..491H,2024ATel16502....1A} , thus additional clouds are needed in our model.
As for the flare before 700 days, apart from the sudden change of the peak frequency at the 4th observation, the complex evolution of the light curve (fast - slow - fast decay) is not predicted by our model (see  Figure \ref{fig:lc_sample}).
Thus, we suspect that the flare before 700 days comes from the interaction of the outflow with at least two closely located clouds.
Overall, we consider these multiple late-time radio flares are from the interaction of the outflow with at least five giant clouds at different distances.
Figure \ref{fig:15oi} shows the modeling of the light curve, and Figure \ref{fig:15oi_s} shows the spectra comparison between the observed  and the predicted. 
The adopted parameter values for these five collisions are shown in Table \ref{tab:pam}.

Here among these collisions, the radial width $t_{w}(r)$ increases with distance due to the radial expansion of the outflow (see Section \ref{sec_outflow}).
After each previous cloud collision, taking into account that part of the outflow's  kinetic energy is dissipated,  we adopt a reduced value of $v_w$ for the next collision.
In addition, as the outflow moves outwards and the density of the outflow decreases, we found that much higher $\epsilon_e$ and $\epsilon_b$ are needed for the last two clouds than the inner clouds to account for the later but brighter flares.

The adopted  velocities ($v_w\sim$0.5c) for ASASSN-15oi and AT2018hyz are  higher than other events because of their brighter luminosities.
Simulations \citep{2018ApJ...859L..20D, 2023MNRAS.523.4136B} show that the outflow has a wide range of $v_w$ about $0.1\sim0.8$ c, depending sensitively on the inclination angle.
Thus, our values are in a reasonable regime.

\section{Discussion}

Here we discuss an off-axis jet, which can produce the steep rise as observed and is proposed to explain these late radio flares \citep{2023MNRAS.522.4565M,2024MNRAS.527.7672S}. 
We also consider the possible accompanying X-ray emission from the outflow-cloud interaction \citep{2021ApJ...908..197M,2021JHEAp..32...11Z}, and compare it with observations.

\subsection{An off-axis jet}

A TDE might generate a relativistic jet \citep{1976MNRAS.176..633F,2012ApJ...760..103D,2012MNRAS.420.3528M}.
The shock interaction between the relativistic jet and the CNM powers non-thermal radio synchrotron radiation \citep{2011MNRAS.416.2102G}.
The radio signal is beamed away from the observers initially if we are not located within the initial jet aperture $\theta_{j}$ (an off-axis jet).
As the jet decelerates,  the emission from the jet becomes detectable to observers at larger viewing angles $\theta_{obs}$, producing a steep rise as $f_{\nu}\propto t^{(15-3p)/2}$  for $\nu> \mathrm{max}(\nu_m,\nu_a)$ \citep{1999ApJ...525..737R,2002ApJ...570L..61G}.

In the off-axis jet scenario, the radio would peak at a time $t_{p}$ with a peak luminosity $\nu L_{\nu}$  ( $\nu$ = 1 GHz) as \citep{2002ApJ...579..699N,2019MNRAS.488.2405G}:
\begin{align}
t_{p}& = 60 \  n_{0}^{-1/3} E_{52}^{1/3}(\frac{\theta_{obs}-\theta_{j}}{15^{\circ}})^{2} [\mathrm{day}]\\
\nu L_{\nu}& \approx 1.1\times 10^{38}\   \epsilon_{e,-1}^{p-1} \epsilon_{b,-3}^{(p+1)/4}n_{0}^{(p+1)/4} E_{52}(\frac{\theta_{obs}}{20^{\circ}})^{-2p}\notag \\
& [\mathrm{erg/s}]
\end{align}
where $E_{52}$ is the energy of the jet in units of $10^{52}$ erg; $n_0$ is the uniform ambient density in $\mathrm{cm^{-3}}$.
The equations reveal that a flare with a peak time of about 1,000 days and a peak luminosity of $10^{39}$ erg/s is possible.

\cite{2016ApJ...816...20L} find that the  the data before 1,000 days in IGR J12580+0134 are consistent with an off-axis relativistic jet.
Besides, \cite{2024MNRAS.527.7672S} also find that the off-axis jet model can explain the full set of radio observations of AT2018hyz.
However, the delayed radio flare seen in ASASSN-15oi cannot be explained by such a model, as \cite{2021NatAs...5..491H} find that its best-fit parameters vary substantially between each epoch and no numerical solution can account for both the initial steep flux rise, and the complex spectral and temporal evolution. 
\cite{2024arXiv240413326S} find that a structured two-component jet (a relativistic inner component and trans-relativistic outer component), instead of the one-component jet, can explain the radio data of ASASSN-15oi before 1,400 days, as well as ASASSN-1eae, AT2018hyz and AT2019dsg.
Nevertheless, the latest observed flux variation after 1,400 days in ASASSN-15oi is unlikely to be explained by this model.

\subsection{X-ray  emission from the outflow-cloud interaction}\label{X_ray}


The interaction of outflow with a cloud can also drive a cloud shock that propagates into the cloud \citep{1975ApJ...195..715M}.
The cloud shock's velocity is $\beta_{sc} =\beta_{w}\chi^{-1/2} \approx 0.003\beta_{w,-1} \chi^{-1/2}_3$, where $\chi_3 = (n_{c}/n_w)/10^{3}$ is the density ratio and $n_{c}$ is the cloud's gas number density.
The shocked cloud material's temperature is $T_{sc} = 3m_{H}c^{2}\beta_{sc}^{2}/16k\approx 6\times10^{6} \beta_{w,-1}^{2}\chi_3^{-1}$ K, which means that the shocked cloud radiates in X-rays.
Below, we estimate its luminosity.

The internal energy gained by the shocked cloud per unit time is $\dot{E}_{\mathrm{c,th}}=\frac{9}{32}\rho_{c}r^{2}v_{sc}^{3}\Omega_c$, and the energy conversion efficiency is $\eta = \dot{E}_{\mathrm{c,th}}/L_w\approx 2\% \frac{\Omega_{c}}{\Omega_w}\chi^{-1/2}_{3}$ where the $L_{w}$ is the kinetic luminosity of the outflow \citep{2021MNRAS.507.1684M}. 
In addition, \cite{2023MNRAS.518.5163C} found that the thermal conduction inside the cloud can play a crucial role in increasing the cloud’s radiation, and up to 5$\sim$10$\%$ of the kinetic energy of the outflow impacting the cloud could be converted into radiation.
In this case, one has $\eta=5\sim10\% \Omega_c/\Omega_w$.

This energy will be radiated out.
The radiative cooling timescale is $t_{cool}= 0.69kT_{sc}/n_c \Lambda (T_{sc})
\approx 5\times10^{6} T_{sc,7}^{1/2}n_{c,7}^{-1}$ s where $T_{sc,7}\equiv T_{sc}/10^{7}$ K, $n_{c,7}\equiv n_{c}/10^{7}$ $\mathrm{cm^{-3}}$ and $\Lambda(T)\approx 6\times 10^{-27}T^{\frac{1}{2}}$ [erg $\mathrm{cm^{3}}$/s]  is the temperature-dependent volume cooling function 
\citep{1976ApJ...204..290R,1998ApJ...500..342B,2000ApJ...540..170L}.

For $t_{cool}<t_w$, the cooling is so strong that the internal energy of the shocked cloud is rapidly lost in radiation. Then, the X-ray luminosity is $\eta L_w(t)$
which would peak at $t^{\prime}=t_{w}(r)$ with a peak luminosity of 
\begin{align}
L_{X} \approx 10^{41}  \Omega_{c,-1}\Omega_w^{-1} n_{w,4}^{1/2} n_{c,7}^{-1/2}L_{w,44}\ \mathrm{erg/s} 
\label{eqn_L_x_1}
\end{align}
where $L_{w,44}\equiv L_w/10^{44}$ erg/s.

For  $t_{cool}>t_w$,  the  peak luminosity can be estimated as $L_X=\eta E_{w}/t_{cool}$ or
\begin{align}
L_{X}&\approx 10^{40}E_{w,50} \Omega_{c,-1} n_{c,7}^{7/22} n_{w,4}^{1/2} \beta_{w,-1}^{-3/11}\notag \\
&\dot{m}_{0,-1}^{-3/11}\Omega_w^{-17/11} R_{in,-1}^{6/11} t_{w,2}^{-5/11}\   [\mathrm{erg/s}]
\label{eqn_L_x_2}
\end{align}
at a peak time $t^{\prime}=t_{cool}$ or
\begin{align}
t_{cool}&\approx 100 n_{c,7}^{-9/11} \dot{m}_{0,-1}^{3/11}\Omega_w^{6/11} \notag\\
&R_{in,-1}^{-6/11} \beta_{w,-1}^{3/11}t_{w,2}^{5/11}\ [\mathrm{day}]
\label{eqn_t_cool}
\end{align}
where $E_{w}=E_{w,50}\times 10^{50}$ erg is the total energy of the outflow.

\subsubsection{Comparison with observations}

\begin{figure}
\centering
\includegraphics[scale=0.35]{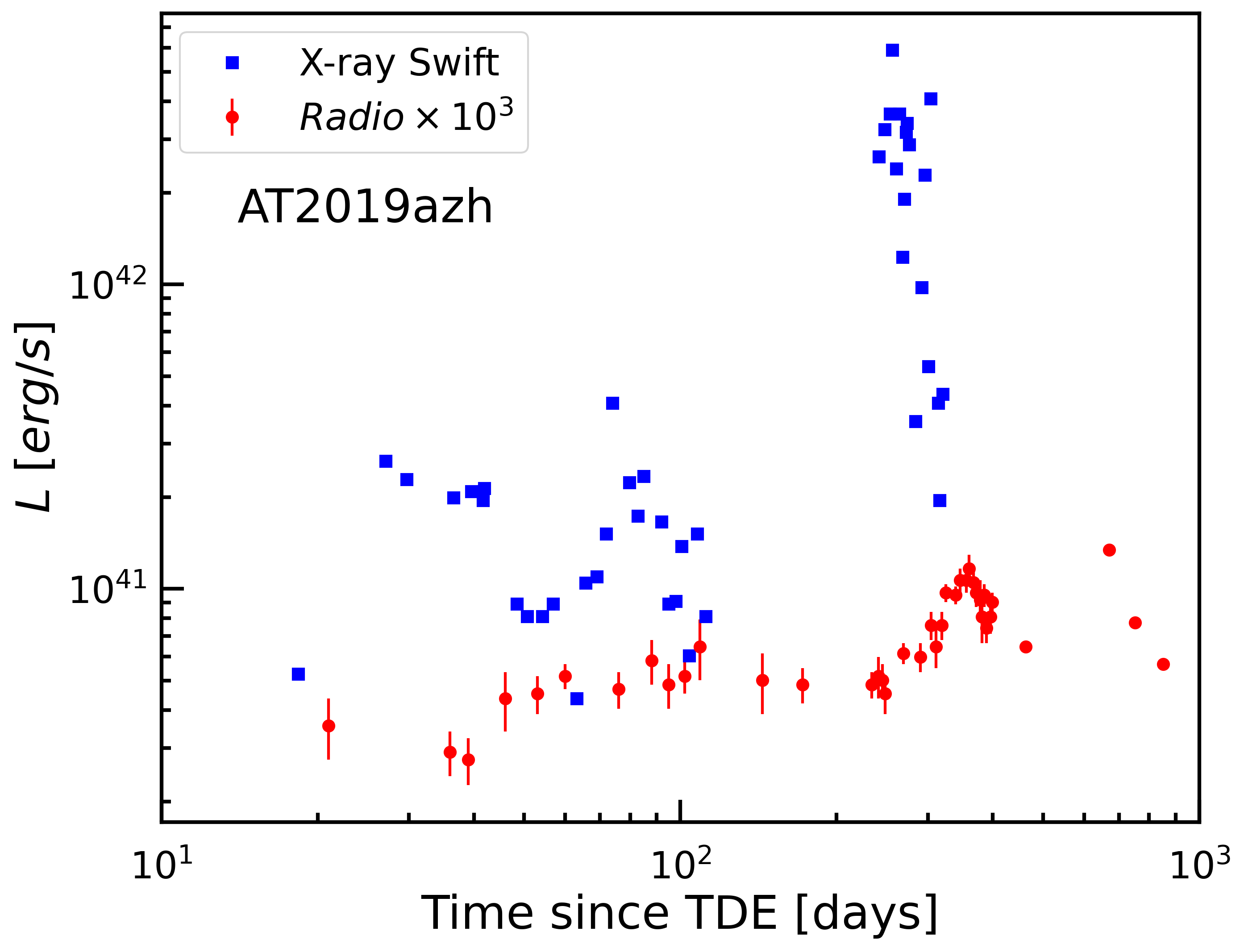}
\caption{The X-ray and radio light curves for  AT2019azh. The X-ray data is from \cite{2021MNRAS.500.1673H}.}
\label{fig_L_x_19}
\end{figure}

\begin{figure}
\centering
\includegraphics[scale=0.35]{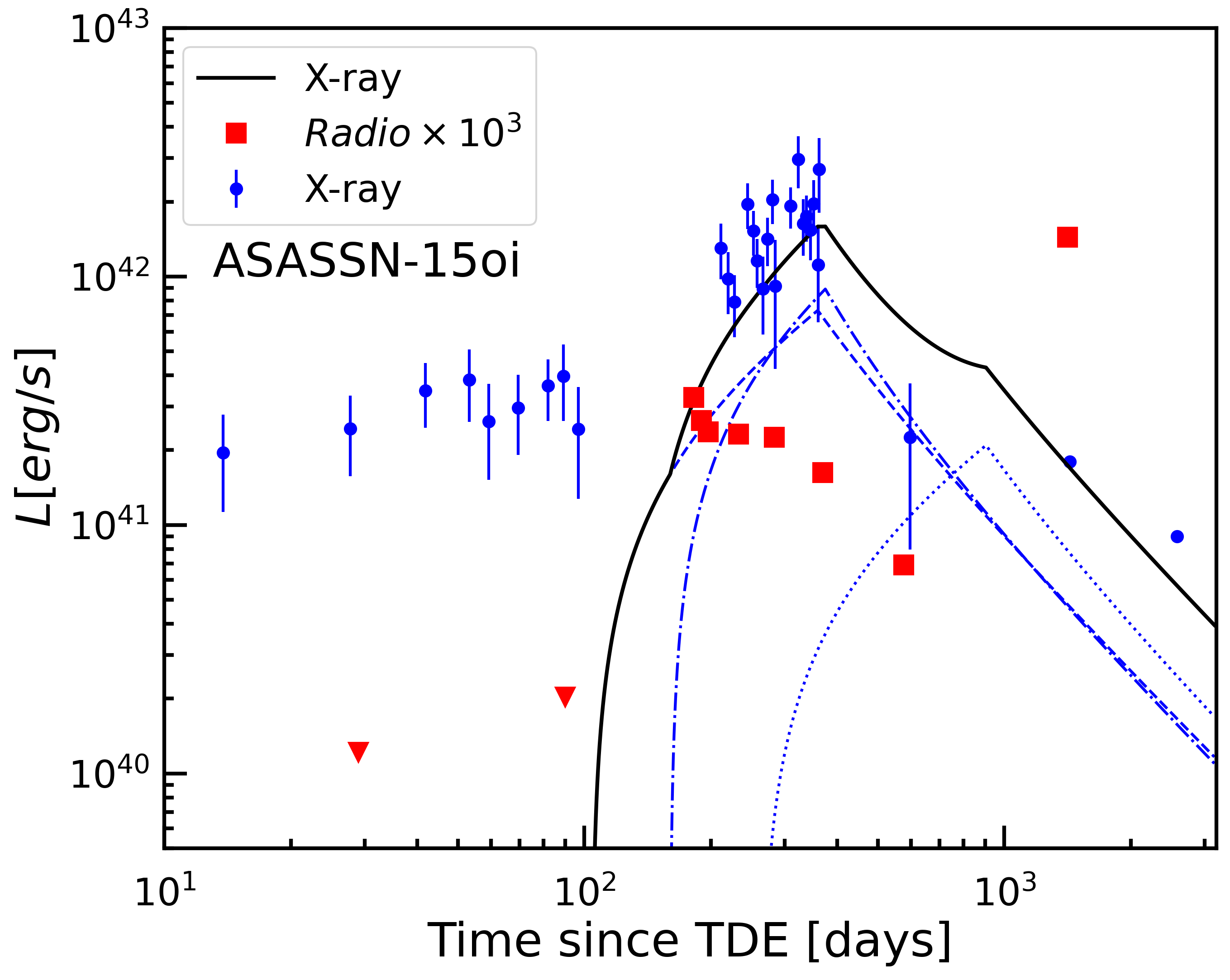}
\caption{The X-ray and radio light curves for ASASSN-15oi. The X-ray data is from \cite{2017ApJ...851L..47G} and \cite{2024arXiv240719019H}. The dashed, dotted-dashed and dotted lines are X-ray emission from the interaction of the outflow with the 1st, 2nd, and 3rd clouds, respectively, while the black line is the total emission.}
\label{fig_L_x_15}
\end{figure}

Below we check the above with the events that show possible X-ray flares accompanying the late-time radio flares.
We found no report of X-ray observation for IGR J12580+0134 and  AT2020vwl near the times of their late radio flares, while there is only one observational X-ray data for AT2018hyz \citep{2022ApJ...938...28C}. 
For AT2019azh \citep{2021MNRAS.500.1673H} and ASASSN-15oi \citep{2017ApJ...851L..47G,2024arXiv240719019H}, both have sufficient X-ray data detected accompanying the radio flare, and their X-ray light curves are shown in Figure \ref{fig_L_x_19} and Figure \ref{fig_L_x_15}, respectively.

For AT2018hyz,  its X-ray luminosity remains nearly constant at $10^{41}$ erg/s up to 250 days \citep{2020MNRAS.497.1925G,2020ApJ...903...31H}.
The latest observation at $t=1253$ days by \cite{2022ApJ...938...28C} shows a luminosity of $5\times10^{40}$ erg/s. 
Our late radio flare modeling for AT2018hyz has found that the collision occurred at $t=800$ days, with $t_w(R_{in}) \sim700$ days.
Then with all those parameters and using Eq. (\ref{eqn_L_x_1}), we found that $n_c=10^{8}$ $\mathrm{cm^{-3}}$ (corresponding to a cloud mass of $\sim10^{5} \ \mathrm{M_{\odot}}$) can produce an X-ray flare in the $t_{w}>t_{cool}$ regime with  $L_X\sim 7\times10^{40}$ erg/s at $t=1500$ days.
The X-ray luminosity from the shocked cloud is still rising at  $t=$1250 day, hence is consistent with the observation.

For AT2019azh, its late-time X-ray shows a peak luminosity of $L_X \approx10^{42}$ erg/s and it peaks earlier than the radio at least 100 days ahead.
However,  according to Eqs. (\ref{eqn_L_x_1}-\ref{eqn_t_cool}), the predicted X-ray peak time is $\approx$max($t_w,t_{cool}$), which is unlikely earlier than the radio that peaks at $t_w$.
On the other hand, we found that adopting  $n_c=10^{7}$ $\mathrm{cm^{-3}}$ will result in an X-ray flare in the $t_{w}>t_{cool}$ regime that peaks at 350 days with $L_X\approx 10^{40}$ erg/s, much lower than the observed.
Thus, the late X-ray flare of AT2019azh may not come from the outflow-cloud interaction and it  may be related to the on-going SMBH accretion instead.

For ASASSN-15oi,  the late-time accompanying X-ray peaks  at about 350 days with  $L_X\approx  10^{42}$ erg/s, and then shows a slow decline to $\sim10^{41}$ erg/s.
The X-ray peak time is $\sim300$ days later than the radio.
Our modeling  has found that three clouds are required for the radio emission before 600 days.
We calculate the X-ray light curves from the collisions of the outflow with each cloud, simply adopting a linear rise and exponential decay light curve shape. We found that the observed X-ray may be accounted for by summing up the emission from the three collisions (see the solid black line in Figure \ref{fig_L_x_15}).
Here, $n_c=10^{6.8}$ $\mathrm{cm^{-3}}$ is adopted for all the three clouds, corresponding to cloud masses of $1$, $20$ and $30$ $\mathrm{M_{\odot}}$, respectively.

\section{Conclusion}

In recent years, radio observation of some TDEs reveals flares only at a late time ($10^{2}\sim10^{3}$  days after TDE discovery), such as in ASASSN-15oi and AT2018hyz, both of which have a radio luminosity of about $10^{39}$ erg/s.
Due to the rapid evolution of light curves, some of these late-time radio flares are difficult to be explained by the conventional outflow-CNM interaction, of which the steepest evolution is $f_{\nu}\propto t^{3}$ \citep{2012MNRAS.420.3528M,2016ApJ...827..127K,2020SSRv..216...81A}  while observation requires a temporal power-law steeper than $t^{4}$.

A variety of explanations have been proposed for these late-time radio flares such as  a delayed launch of the outflow \citep{2023arXiv230813595C}, a misaligned precessing jet \citep{2023ApJ...957L...9T}, 
an off-axis jet \citep{2023MNRAS.522.4565M,2024MNRAS.527.7672S},
and a piecewise power-law distribution of CNM \citep{2024arXiv240415966M}.

In this paper, we developed the outflow-cloud interaction model for the late-time radio flares.
This collision forms a bow shock, and the late-time radio synchrotron emission is produced from the shocked outflow.
We found it is capable of generating bright radio flares years after TDE with a luminosity of $10^{39}$ erg/s, and can explain the large delay,  the sharpness of the rise and the multiplicity of the late radio flares.

The model is applied to five TDE candidates and our model can well explain the observed data, including the multiple late-time flares such as seen in ASASSN-15oi, when the outflow impacts multiple clouds.
The inferred outflow velocity ranges from 0.2$c$ to 0.6$c$ with a mass outflow rate of about $0.01\sim0.1\ \mathrm{M_{\odot}/year}$.
The corresponding kinetic luminosity of these outflows is about $10^{44\sim45}$ erg/s which is consistent with the simulation \citep{2019MNRAS.483..565C,2023MNRAS.523.4136B,2023MNRAS.521.4180B}, and the total kinetic energy of the outflow is  $10^{50}\sim10^{51}$ erg with a total mass of about $0.01\sim0.1 \ \mathrm{M_{\odot}}$.
The distances  ($\sim0.1$ pc) and  the number of  clouds required are compatible with observations \citep{1996A&ARv...7..289M,2005ApJ...622..346C}.

Our model may be tested by the accompanied X-ray or dust echo emission component.
Here in this paper, with the parameters obtained by the radio modeling, we found that the X-ray emission in ASASSN-15oi and AT2018hyz can be explained by the shocked clouds, while the X-ray in AT2019azh may come directly from some on-going accretion.
In addition, there may be lots of dust in the torus as well, such that the UV emission from the TDE's early accretion may be absorbed and reprocessed into the infrared \citep{2016MNRAS.458..575L}.

With more late-time radio flares by future observation, we will examine whether our model can be applied to these events and offer constraints on the outflow to help better understand TDEs. In addition, they can also be used as a tool to study the circumnuclear environment.

\section*{ACKNOWLEDGEMENTS}
We thank the anonymous referee for many constructive comments and suggestions that helped improve the quality of this paper.
We thank Adelle Goodwin for helpful comments.
RFS and JZ are supported by National Natural Science Foundation of China (grants 12073091, 12393814 and 12261141691) and by the Strategic Priority Research Program of Chinese Academy of Sciences (grant XDB0550200). GM is supported by the NSFC (No.12473013, 12133007).


\bibliography{sample631}{}
\bibliographystyle{aasjournal}

\end{document}